\documentclass[]{aastex631}

\usepackage{makecell}
\usepackage{multirow}
\usepackage{threeparttable}
\usepackage{amsmath}
\usepackage{graphicx}

\shorttitle{Modeling FETCH CME}
\shortauthors{Jensen et al.}

\begin{document}
\title{Modeling Polarized Radio Sounding Observations of a Coronal Mass Ejection}

\author[0000-0003-0216-6621]{Elizabeth A. Jensen}
\affiliation{Planetary Science Institute \\
1700 East Fort Lowell, Suite 106 \\
Tucson, AZ 85719-2395}
\altaffiliation{ACS Engineering \& Safety, Spring, TX 77373}

\author{W. B. Manchester IV}
\affil{Climate and Space Sciences and Engineering, University of Michigan, Ann Arbor, MI 48109, USA}

\author{J. E. Kooi}
\affil{Remote Sensing Division, U.S. Naval Research Laboratory, 4555 Overlook Ave. SW, Washington, DC 20375, USA}

\author{T. Nieves-Chinchilla}
\affil{Heliophysics Science Division, NASA Goddard Space Flight Center, 8800 Greenbelt Rd., Greenbelt, MD 20771, USA}

\author{L. K. Jian}
\affil{Heliophysics Science Division, NASA Goddard Space Flight Center, 8800 Greenbelt Rd., Greenbelt, MD 20771, USA}

\author{D. B. Wexler}
\affil{Space Science Laboratory, University of Massachusetts Lowell, 600 Suffolk St., Lowell, MA 01854, USA}

\author{S. F. Fung}
\affil{Heliophysics Science Division, NASA Goddard Space Flight Center, 8800 Greenbelt Rd., Greenbelt, MD 20771, USA}

\author{N. Gopalswamy}
\affil{Heliophysics Science Division, NASA Goddard Space Flight Center, 8800 Greenbelt Rd., Greenbelt, MD 20771, USA}

\begin{abstract}

Coronal Mass Ejections (CMEs) evolve significantly as they propagate from the Sun to the Earth, so remote observations of their changes in speed, strength of the magnetic field, density, and overall structure are critical for predicting their arrival time and geoeffectiveness. Radio line-of-sight observations of Faraday rotation and Total Electron Content combined with white-light observations enables the measurement of these properties with careful analyses. This paper describes the analysis techniques and evaluates their accuracy with regard to measuring a CME’s complex evolving structure and speed. The approach utilizes the layout of the Faraday effect tracker of coronal and heliospheric structures (FETCH), a purely space-based instrument concept, with Alfv\'en Wave Solar atmosphere Model (AWSoM) simulations as input for evaluating these radio-based measures. Focusing on density and velocity/speed, we find that {\em in-situ} measurements of CME properties observe similar but different aspects of the distinct CME structure. The AWSoM model suggests that Faraday rotation may be a more sensitive measure of structure than Total Electron Content (TEC). Finally, we discuss the difficulty the simulation reveals in determining the trailing edge location of a magnetic flux rope.

\end{abstract}


\section{Introduction}\label{sec:intro}

It is very well established that Coronal mass ejections (CMEs) in their interplanetary forms (ICMEs) are the main drivers of harmful space weather such as ionospheric disturbances that undermine communication signal fidelity and GPS accuracy, while increasing satellite drag and the risk of electrical power failures by induced currents \citep{Gombosi:2017}.  These large-scale disturbances in the heliosphere involve a sudden release of massive amounts of solar plasma often containing intense magnetic fields, which can accelerate to high speeds. The internal magnetic strength and topology within an ICME can trigger a geomagnetic storm \citep{Gosling:1993a}, while the shock at the ICME leading edge can accelerate particles to very high energies (e.g. \citealp{Axford:1972, Lee:1997, Reames:1999, Zank:2000, Giacalone:2002, Li:2003,  Sokolov:2004, Roussev:2004, Tylka:2005, Zhao:2024}). The evolving geomagnetic storms occurring after the impact of an ICME and shock with the magnetosphere, and the associated energetic particle events can adversely affect people working in the upper reaches of Earth's atmosphere and equipment all the way from orbit to the bottom of the ocean (e.g. \citealp{2022Atmos..13.1781G}). Understanding and determining the structure of these CMEs/ICMEs and their evolution in the heliosphere is essential for predicting and preparing for space weather. 

CME/ICMEs have been observed in the heliosphere with a number of wide-angle heliospheric imagers including the Solar Mass Ejection Imager (SMEI) \citep{Eyles:2003, Jackson:2004} and the Heliographic Imagers HI1 and HI2 (part of the Sun Earth Connection Corona and Heliospheric Imager on board the Solar Terrestrial Relations Observatory ``STEREO" dual spacecraft \citep{Howard:2008}).  The valuable remote sensing provided by these instruments is augmented by in situ observations from instruments onboard spacecraft such as the Advanced Composition Explorer (ACE) \citep{Stone:1998}, Wind, and STEREO A and B \citep{Kaiser:2008}.  These measurements include the magnetic field strength \citep{Lepping:1995} and plasma quantities \citep{Ogilvie:1995}, which provide localized in-sights of the internal structure of ICMEs.  When combined, remote and {\em in-situ} observations allow for comprehensive multi-spacecraft analysis of CME/ICME events (e.g. \citealp{Rouillard:2009,Mostl:2010,Kilpua:2011,Ruffenach:2012,Janvier:2015,Kilpua:2019,Davies:2022}).  Recent increases in observational capability come with the ground-based (e.g. Daniel K. Inouye Solar Telescope, \citealp{2021SoPh..296...70R}) and the space-based observatories Parker Solar Probe and Solar Orbiter (\citealp{2010AcAau..67.1063G, 2020A&A...642A...1M}). These studies have enabled significant progress in understanding the 3D evolution of CMEs/ICMEs \citep{Kilpua:2017}  and the physical processes responsible for CME/ICME evolution \citep{Manchester:2017}. 

Even with these significant advances there remains a scarcity of remote observations that prevents us from fully characterizing CMEs/ICMEs.  However as we will show, in many ways Faraday rotation (FR) observations provide a wealth of remote observations that can inform us of the global morphology, the internal magnetic topology, and the evolutionary changes as the structure approaches the Earth. CMEs/ICMEs exhibit both the size and intensity which can greatly influence radio wave transmission to Earth from sources occulted by the solar atmosphere. Consequently, CMEs/ICMEs introduce numerous propagation effects to trans-coronal radio signals in the form of interplanetary scintillations (see \citealp{2023SoPh..298...22T,2023SpWea..2103396M}) and FR that can be detected by radio telescopes, enabling remote sensing of these exceptional events (e.g. \citealp{Kooi:2022,jensen_2024_10987687}). FR observations have detected clear magnetic structure in several CME events (discussed in the next paragraph). Numerical simulations of FR measure have demonstrated that for idealized models, the magnetic field orientation and helicity of a flux rope can be determined 2–3 days before it reaches 1 au, providing a valuable space weather forecast. FR observations can also resolve the portion of CME/ICME flux ropes that curves back to the Sun \citep{Liu:2007}. 

To date, observing CMEs/ICMEs with FR observations is relatively rare. The first FR observations of the turbulent solar corona that we are aware of were conducted in 1968 using Pioneer 6 as a source \citep{Levy:1969}. An unusual change in the observed FR curve was observed during this superior conjunction consisting of `W'-shaped changes in FR with rotation angles of 40 degrees recorded over a period of 2–3 hr. These `W' features have been attributed to a coronal streamer stalk \citep{Woo:1997} or the passage of a series of CMEs \citep{Patzold:1998}. \cite{Jensen:2008} modeled these structures as CMEs; however, as discovered in \cite{Jensen:2018}, these structures share the jagged characteristics of likely reconnection regions (discussed in the last section of this paper). Later, unusual FR structures contained within CMEs were observed with Helios spacecraft (October/November 1979). White light coronagraph data confirmed the cause of these structures to be CMEs; however, none showed the `W' characteristic \citep{Jensen:2008}.

Advancements in ground-based observation technology have facilitated the detection of FR associated with CMEs/ICMEs by utilizing natural radio astronomical sources, a departure from reliance on artificial satellites in ecliptic orbits. State-of-the-art telescopes, such as the Karl G. Jansky Very Large Array (VLA) and the Robert C. Byrd Green Bank Telescope (GBT), have played a pivotal role in this development. Pioneering studies by \cite{Spangler&Whiting:2009} and \cite{Kooi:2017} successfully identified CME-induced FR within the 1 - 2 GHz frequency range using the VLA. Furthermore, \cite{Howard:2016} detected a subtle CME/ICME FR signal through pulsar observations, albeit at levels comparable to local ionospheric FR signatures only, leading to the establishment of upper bounds on plasma density and magnetic field parameters for the CME/ICME in question. Therefore, the FR methodology has proved to be the possible link between white light imagers and {\em in-situ} observations to connect morphology, internal magnetic topology and changes due to evolutionary processes. 

Interpreting these observations has relied on simplistic solar models for determining the likely distribution of physical quantities required to produce the remotely-measured integrated observation. The knowledge gained from the observations and their analysis has guided the approach taken in this paper, the calculations in section \ref{sec:approach}. Because space weather prediction depends on MHD simulations of ICMEs, these models have advanced into 3D represenations of the heliosphere. \emph{Consequently, these models enable examining FR analysis techniques for how well they measure the ICME structure and flow, keeping in mind the limitations of the model}. For this paper, we have selected the \cite{Manchester:2014a} Alfv{\'e}n Wave Solar Model (AWSoM) model, discussed in section \ref{sec:model}. Remote observations of asymmetrical structures such as ICMEs are strongly dependent on viewing geometry; perpendicular to the flow provides the simplest remote observing. This is why the configuration of the Multiview Observatory for Solar Terrestrial Science (MOST) mission spacecraft is selected for this analysis as described in section \ref{sec:MOST}.

In order to develop a more in-depth study of the suitability of the FR technique to investigate the CMEs/ICMEs, we have divided the analysis in two papers. This initial paper aims to explore various aspects of FR observing through an ICME simulation. Specifically, it will investigate (a) the additional insights that can be gleaned from analyzing FR datasets, (b) the optimal approach for establishing error margins on magnetic fields derived from FR measurements, and (c) the potential advantages of strategically situating FR instruments to observe plasma upstream of Earth. Fitting the FR observations of an ICME with a magnetic flux rope model is out of scope for this paper and will be addressed in a separate paper. This paper performs the initial analyses that need to take place prior to the fit. In order to do a fit, two questions have to be answered first. In this paper, we answer them: (1) How well was the parallel component of the magnetic field in FR separated from the electron density as provided by Total Electron Content (TEC)? (2) Where are the leading and trailing edges to an ICME's magnetic flux rope located based on the simulated TEC time series? Simulated TEC is obtained via the technique shown in \citep{Jensen:2016}.

\subsubsection{Terminology and Definitions}
The following terms are used, which have specific meanings within this paper:
\begin{itemize}
    \item ICME refers to the plasma properties of the Interplanetary counterpart of a CME. Beyond 20 solar radii, CMEs are generally called ICMEs.
    \item Lines: These are the signal paths between MOST spacecraft for the FETCH instrument shown in Figure \ref{fig:MOST}, labeled 1-4. 
    \item Line-of-sight (LOS): This is a more generic term for signal paths, not those specifically for the FETCH instrument.
    \item Offset: This is the distance from the Sun to a point on the line-of-sight.
    \item Point of Closest Approach: This is the offset point that is at a minimum distance from the Sun.
    \item Impact Parameter: This is the distance from the Sun to the point of closest approach.
    \item GSE Coordinates (Section \ref{sec:synthdata}): Geocentric solar ecliptic coordinates set the x-direction from the center of the Earth to the Sun, the z-direction is to the north out of the ecliptic, and the y-direction completes the right-handed coordinate system.
    \item HGI Coordinates (Section \ref{sec:anothersubsection}): Heliographic Inertial coordinates are sun-centered with the x-axis projecting where the ecliptic and equatorial plane of the Sun intersect, also known as the vernal equinox. The z-axis is the Sun's rotational axis, and the y-axis completes the right-handed coordinate system. 
\end{itemize}

\section{The FR technique on space missions} 

\subsection{Physical Principles of the FR technique} \label{sec:approach}

Analysis of linearly polarized signals provides information on the LOS-integrated product of electron number density ($n_e$) and LOS-aligned component of the magnetic field ($\overrightarrow{B}\cdot d\overrightarrow{S}$). The equation below  is specifically for FR in the regime where the radio frequency ($f_0$) is well above the electron plasma frequency.

\begin{equation}
\begin{array}{c}
    FR = \frac{A}{f_0^2}\int n_e\overrightarrow{B}\cdot d\overrightarrow{S} = \frac{\varphi_{rcp}-\varphi_{lcp}}{2} \\
    RM = \frac{FR}{\lambda^2}
\end{array}
\end{equation}

\noindent with $A=2.36\times 10^4$ rad m$^2$/(T s$^{2}$), $n_e$ electron density in m$^{-3}$, B magnetic field in T, and S distance along the path in m. A convention that we commonly use to express FR is to normalize it by wavelength $\lambda^2$ to calculate the Rotation Measure (RM). The difference in the right-handed polarized (rcp) and left-handed polarized (lcp) phases of the arriving radio wave $\varphi$ is the FR effect.

Frequency fluctuations (FF) consist of shifts in frequency that occur on the carrier frequency of the signal due to {\em changes} in the TEC ($FF = \Delta\omega$, where $\omega = 2\pi f_0$). This is a frequency-dependent measurement obtained using Equation \ref{eqn:ff} adapted from \cite{Jensen:2016}. Integrating the change in TEC enables measuring TEC in time with the radio observing; however, the initial TEC is unknown and the use of white light electron density data is expected to provide this information. Figure \ref{fig:ffcomp} shows the typical sort of FF data that could be expected with one significant modification: the AWSoM output was too smooth, so it was `roughened' up as described in section \ref{sec:anothersubsection}.

\begin{equation}
    \label{eqn:ff}
\begin{array}{c}
    \frac{\Delta \omega}{\omega} = \frac{q^2}{2\omega^2cm_e\epsilon_0}\frac{d}{dt}TEC       \\
    \text{where}\\
      TEC = \int{NdS}
\end{array}
\end{equation}

The fundamental challenge with FR and TEC observing is the ability to decouple the parallel magnetic field and the density and their distribution along the LOS. The first order approach is to assume that they are evenly distributed and to calculate $<B_{\mathrm{\parallel}}> \approx FR / TEC \times f_0^2 / A$. As shown in Figures \ref{fig:Line1} through \ref{fig:Line4} Panels R3, this has an average error around a factor of 2. Using models and/or observing conditions which enable tomographic calculations, the decoupling can be more physically relevant.

\subsection{Veteran Missions Legacies}

Since the era of Helios, the practice of FR observation has typically been augmented with white-light coronagraph imaging whenever feasible. A notable aspect of the study conducted by \cite{Kooi:2017} lies in its integration of {\em in-situ} spacecraft measurements, an extremely rare ``ground truth" moment, of the observed ICME. The scarcity of CME/ICME observations via FR can be attributed to a combination of factors including a lack of fortuity and opportunities. Recent missions, including Magellan, Cassini, MESSENGER, STEREO, Hayabusa, and Mars Reconnaissance Orbiter, have contributed to the accumulation of FR data. Among these endeavors, MESSENGER stands out as the sole mission to have successfully measured FR through a CME \citep{Jensen:2018}.

The MESSENGER CME observation conducted by \cite{Jensen:2018} displayed minimal to negligible presence of a sheath region as it traversed beyond 4 solar radii. Conversely, investigations of CMEs/ICMEs captured by Helios \citep{Patzold:1998}, VLA \citep{Kooi:2017}, and Low-Frequency Array (LOFAR) (M. Bisi, personal communication) suggest that the formation of the sheath region occurs at greater distances from the Sun. Based on white-light data, sheath formation occurs for some ICMEs, but not all \citep{2012ApJ...744...72G,2013AdSpR..51.1981G}. All of these observations captured CMEs/ICMEs traveling perpendicular to the Earth-Sun line (not towards the Earth). While this configuration is useful for studying CMEs/ICMEs in general, the Earth's ionosphere presents an obstacle to making the observations of the ICMEs structural evolution further from the Sun. The ionosphere's TEC and magnetic field strength over its short columnar distance begins to rival the heliosphere's over its au-sized distances. For example, \cite{Howard:2016} discusses this issue complicating the result.

For observing incoming ICMEs from the Earth, \cite{2010SoPh..265...31J} and more recently \cite{2023SoPh..298...74J} show that the ICME's sheath also presents an obstacle to studying the structure (looking down the `nose'), reducing the overall FR. Developing the $B_{\perp}$ observing technique \citep{2024ApJ...963...25J} enables distinguishing the flux rope from the sheath due to its poloidal curvature could potentially make this approach more viable but only for a single estimate per source. Much of the CME/ICME evolution takes place when the ionospheric effects become too significant.

One of the major motivations for focusing on CMEs/ICMEs with FR observations is to obtain a remote measure of the strength and orientation of the magnetic field in the flux rope within. Ever since the prevalence of poloidal/torroidal structure of CME/ICME magnetic flux ropes was discovered, attempts to interpret FR observations using these structures have been undertaken (for example, \citealp{Liu:2007} and \citealp{2020ApJ...896...99W}). The clearest and most successful observation of the flux rope structure in a CME with FR was in \cite{Jensen:2018}.

\begin{figure}
    \centering
    \includegraphics[width=6in]{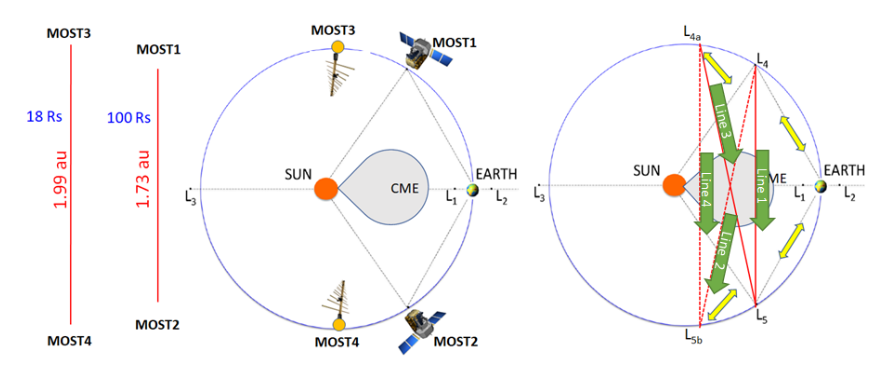}
    \caption{Overview of the MOST mission with the four constituent spacecraft at L4 (MOST1), L5 (MOST2), ahead of L4 (L4a, MOST3) and behind L5 (L5b, MOST4). MOST3~\&~4 will carry only radio equipment for Faraday rotation measurements (FETCH). MOST1~\&~2 will have identical remote-sensing and in-situ instrument suites. The approximate MOST1-MOST2 and MOST3-MOST4 distances are shown at the left indicating the long signal paths for spacecraft radio signals (blue text shows the offset distance to the point-of-closest-approach between the LOS and the Sun). The red lines in the right indicate FETCH signal paths; the green arrows label these paths Line1-4 and give the forward directions; this matters for the sign of Faraday rotation. This enables a rudimentary radio tomographic image using TEC. The yellow double arrows indicate communication links to Earth. Note that MOST3~\&~4 slowly shift position throughout the MOST mission. [adapted from \cite{2024JASTP.25406165G}]}
    \label{fig:MOST}
\end{figure}

\subsection{MOST/FETCH: a new mission concept} \label{sec:MOST}

The proposed Multiview Observatory for Solar Terrestrial Science (MOST) mission as presented to the 2024-2033 Solar and Space Physics Decadal Survey seeks to understand the magnetic coupling of the solar interior to the heliosphere (\citealp{2024JASTP.25406165G,2023FrASS..1064069J}). Composed of 4 spacecraft, 2 hosting 10 instruments in the Lagrange L4 and L5 positions and 2 hosting the Faraday Effect Tracker of Coronal and Heliospheric Structures (FETCH) instrument alone at larger angular offsets from Earth, MOST will be the most comprehensive solar mission to date. To be clear, FETCH is hosted on all four MOST spacecraft. The locations of its deployed spacecraft are chosen to improve on STEREO's work observing upstream from the Earth and L1, but with new technology (FETCH) capable of remotely observing the magnetic field in the Earthbound solar wind as well using FR. 

The FETCH instrument concept envisages four radio Lines-Of-Sight (LOS) for measuring the electron density, magnetic field parameters, and velocity/speed parameters crossing the LOS as shown in Figure \ref{fig:MOST}. FETCH is a radio sounding instrument, detecting the modifications that a controlled signal undergoes due to the plasma through which it is traversing. The cutting edge technology of FETCH will enable it to be the first instrument to pass signals from spacecraft to spacecraft across 2 au, and with the greatest sensitivity in the signal (the lowest offset point) occurring where the Earth-bound solar wind plasma is passing the LOS. As we showed above, the signal characteristics of frequency, phase, and polarization are combined to measure FR, TEC, and Frequency Fluctuations (FF) using established techniques (e.g. \citealp{Jensen:2018,2017SpWea..15..310W,2000AstL...26..544E,2014ApJ...788..117I}).

\section{Methodology}

Besides the scientific inquiry provided by this manner of observing, this study investigates the opportunities and challenges with using spacecraft polarized radio signals to observe ICMEs. We simulated the FR, TEC, and velocity of an Earth-directed ICME crossing the FETCH radio sensing paths. The necessary simulation data, magnetic field, density, and velocity, along each FETCH LOS was obtained from the AWSoM heliospheric 3-D MHD model. We inspected the parallel magnetic field, speed, expansion speed, magnetic flux rope region, sheath, and leading/trailing edges in how they compare between the MHD model's output, the simulated FETCH data from the model's output, and what an {\em in-situ} spacecraft would observe within this output solution space. We then acknowledge the limits between this simulated investigation and the actual dynamics in the heliosphere.


\subsection{MHD Modeling} \label{sec:model}

CMEs/ICMEs were discovered through Thomson-scattered white light imaging with with early space-born coronagraphs. Today, modern instruments continue to provide essential observations for space weather research. For instance, the STEREO heliographic imagers facilitated the creation of three-dimensional reconstructions, as highlighted by \cite{Howard:2008}, while the SMEI spacecraft, as discussed by \cite{Eyles:2003}, tracked the propagation of ICMEs well beyond Earth's vicinity. Although white light observations of CMEs/ICMEs have been instrumental in identifying their morphological characteristics, they do not provide magnetic properties, crucial to assess their geoeffectiveness. There are still unanswered questions regarding the correlation between the visual electron density observations and in-situ magnetic field and plasma measurements (\citealp{2002PhDT........31M}). Apart from {\em in-situ} observations, one has to use indirect observations to infer the magnetic structure of CMEs/ICMEs by combining forward modeling, photospheric magnetic field, and post eruption arcade properties \citep{2018IAUS..335..258G,2018JASTP.180...35G}. There is currently not a way to globally measure the magnetic field structure of CMEs/ICMEs. Addressing this limitation is the motivation of this paper and the analysis techniques described below.

In the absence of remote magnetic field measurements accompanying the plasma density imaging, MHD models enable making estimates of this critical characteristic using a variety of physical requirements and mathematical structures (grids, boundary conditions, and scaling methods). MHD provides a fluid description of plasma environment, providing the local properties of the flow including mass density, bulk velocity and temperature along with the magnetic field. In the ideal case, the system of equations are the Euler fluid equations combined with the induction equation describing the time evolution of the magnetic field. The Lorentz force term then enters the momentum equation while magnetic terms enter the energy equation completing the system. Such a system represents the first three moments of the Boltzmann equation including magnetic fields in an equation of state. The MHD equations also can be written as conservation laws for mass, momentum, and energy to solve for the response of heliospheric plasma given initial conditions and boundary conditions. They can vary from 1 dimensional for the simplest to 3 dimensional, the numerical solution grid can remain fixed or adapt to the scale sizes of critical plasma parameters for the fluid solution to remain valid, and the time scales are generally selected for stability. 

In this work, we are using a much more realistic MHD simulation of the ICME than \cite{Liu:2007}, the AWSoM simulation of the 2005 May 13 CME/ICME. In the AWSoM model, the coronal heating and solar wind acceleration are driven with low-frequency Alfv\'en wave turbulence. A simple phenomenological turbulence formulation of wave energy transport with local turbulence dissipation is applied. The Alfv\'en waves emanate  from the bottom boundary and propagate parallel to magnetic field lines. In this original model, wave reflection was parameterized by means of a reflection coefficient that prescribed the ratio of returning and outward propagating waves uniformly. Wave dissipation is driven by the mixing of forward and counter propagating waves, with balanced turbulence, optimal near the tops of closed loops. The dissipation depends on the correlation length $L_\perp$ that we assign inversely proportional to $\sqrt{B}$: $L_\perp\sqrt{B}= 7.5\times10^4$. In the following work, we prescribe 40\% of the turbulently dissipated energy to electron heating, while the remaining 60\% heats the protons.

\subsection{May 13 2005 CME/ICME Simulation: ``Ground Truth" for Investigation}\label{sec:subsubsection}
The May 13 2005 CME erupted from the north-south polarity inversion line of NOAA active region (AR) 10759 at 16:03 UT, reaching speeds around 2000 km/s in the corona. It was selected for this study because the ICME that resulted impacted the Earth's geospace at around 1000 km/s with peak magnetic field (axial) strength of around 60 nT southward and produced a severe geomagnetic storm. 

For this analysis, we needed a relatively realistic model for the ICMEs physical parameters. One issue relative to this study is that caution needs to be taken to make sure that any interpretations are those that are independent of the simulation model used. That the model enabled our timing ICME structural components to obtain speed is a general mathematical result regardless of which model was used. Attempting to place the trailing edge to the magnetic flux rope where there was a local increase in FR is model dependent, for example. Hence, we worked to find another approach for this parameter that would be consistent among the models.

Our coronal model for Carrington rotation 2029 includes AR 10759 from which the CME erupts.  The computational domain is divided into two components, a solar corona (SC), which extends to $24 R_s$ and an inner heliosphere (IH) domain which extends from $22 R_s$ to 2 au.  The SC employs a spherical adaptive grid with extremely high refinement in the transition region $(\delta r \approx 0.001 R_s)$, and much larger cells $(\delta r \approx 0.25 R_s)$ near the outer boundary located at $24 R_s$.  A cone-shaped region of enhanced refinement is centered on the AR 10759 with resolution four times higher than the standard model. The IH component also has enhanced resolution in the path of the CME, with cubic cell sizes varying from 0.4-0.8-1.6 $R_s$ with refinement levels changing at 20 and 80 $R_s$ from the Sun. 

Following \cite{Manchester:2004}, the CME is initiated with a Gibson-Low spheromak flux rope \citep{Gibson:1998}, which is linearly superimposed upon the existing corona so that the mass and magnetic field of the flux rope are added directly to AR 10759. The flux rope axis and its associated  polarity inversion line (PIL) runs north-south, parallel to the observed polarity inversion line of AR 10759. The set up process used in \cite{Manchester:2014a} was later automated to create the Eruptive Event Generator Gibson-Low \citep{Jin:2017b}. The model mimics the classical three-part structure from which CMEs originate \cite{Hundhausen:1993}. The magnetic field strength reaches a maximum value of 40-50 Gauss, which extends high into the corona in a non-potential configuration.  

\begin{figure*}[ht!]
\centering
\includegraphics[width=1.\textwidth]{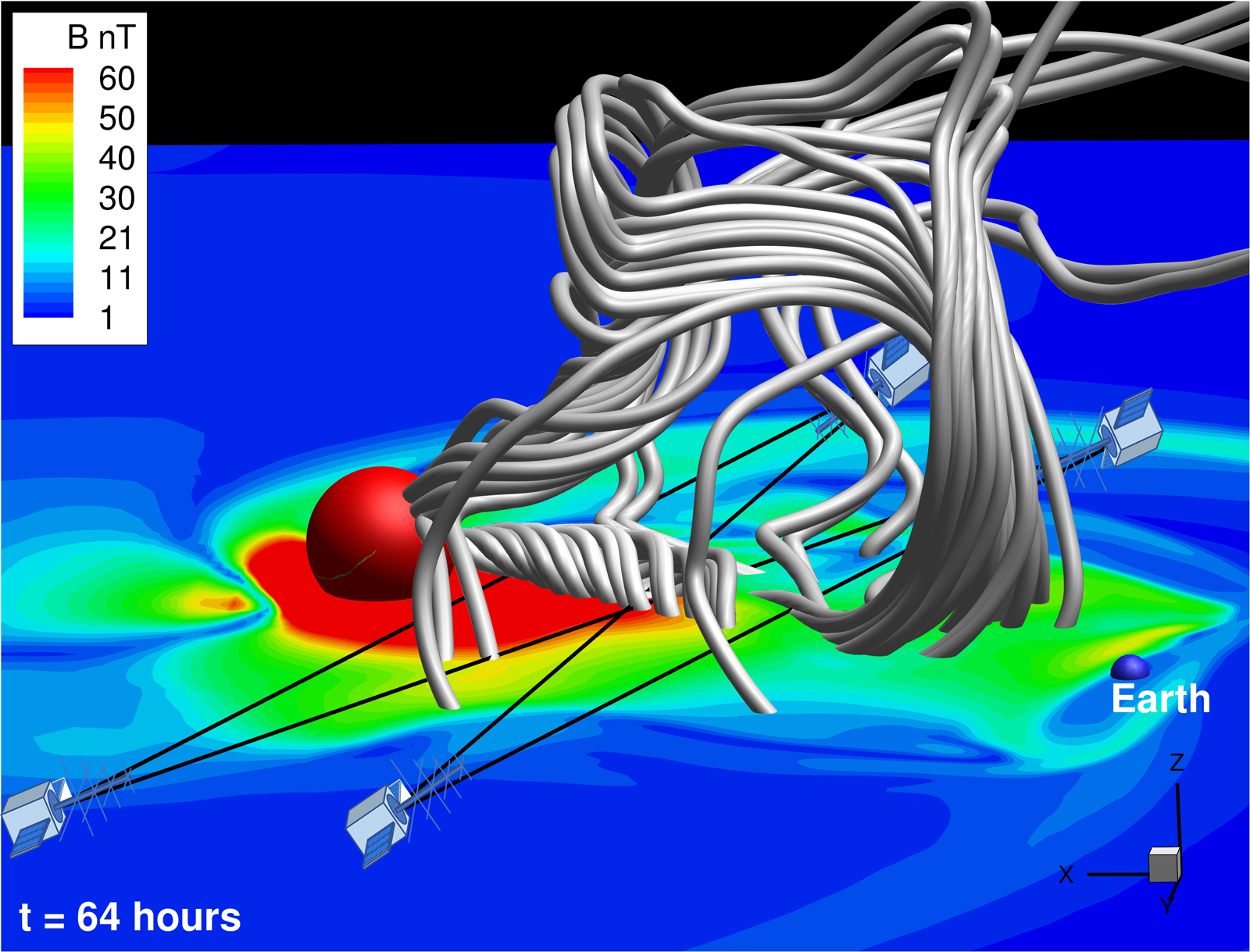}
\caption{3D rendering of the ICME simulation of the May 13, 2005 event 64 hours after initiation. Here, we show the ICME with magnetic field lines in grey. The solar equatorial plane is colored to show the magnetic field strength (not direction). The magnetic field on the 20 Rs source surface sphere is also shown. The field lines emerge from the positive polarity to the north and close in the negative polarity near the equator, making this magnetic flux rope southward oriented. The locations of the MOST spacecraft in the ecliptic plane are shown with the lines of sight (black) running between them.}
\label{fig:CMEmodel}
\end{figure*}

In the low corona, the flux rope rapidly accelerates to speeds of approximately 2000 km/s driving a fast-mode MHD shock ahead of the CME. After 1 hour, the flux rope has reached 15 $R_s$ and its structure resembles a magnetic cloud following self-similar evolution of the initial state.  Eventually, this self-similar behavior breaks down as magnetic reconnection between the flux rope and ambient solar wind sets in. Reconnection at the bottom front and top rear sides opens up the flux rope (relative to the Z-direction), which is followed by a later phase of reconnection at the backside of flux rope.  By 24 hours, the flux rope is turned with its axis nearly perpendicular to the solar equatorial plane as seen in Figure \ref{fig:CMEmodel}. The ICME simulation for this event maintained the magnetic cloud structure with elevated magnetic field strength and reduced temperature and mass density to 2 au. The north-south Bz polarities remain from the original flux rope at 1 au, even though the original rope structure has been greatly modified by reconnection. \citep{Manchester:2014a}

While the ICME speed matches observations one hour after initiation, the model ICME is significantly slower at Earth with a speed of 600 km/s compared to the observed 950 km/s. Here, the excessive deceleration of the simulated ICME is due to the density of the solar wind being roughly a factor of two greater than what is observed, which also increases the FR (RM). The solar wind density has been remedied in more recent models, which show close agreement with observations \citep{vanderHolst:2014awsom, Sachdeva:2019, Sachdeva:2021}.  

For the current study, the four signal paths connecting the four MOST spacecraft (see Figure \ref{fig:MOST}) are created where data is extracted from the AWSoM model's $z=0$ equatorial plane by way of linear interpolation. Along each line, 500 points are uniformly distributed along each line.  Depending on the slope of the four lines relative to the Cartesian axes the step sizes vary as follows: (a) Line 1 $\Delta X=0.25 R_s$ and $\Delta Y=0.71 R_s$, (b) Line 2 $\Delta X=0.41 R_s$ and $\Delta Y=0.70 R_s$, (c) Line 3 $\Delta X=0.12 R_s$ and $\Delta Y=0.81 R_s$, and (d) Line 4 $\Delta X=0.28 R_s$ and $\Delta Y=0.81 R_s$. While the lines are extracted from the model at a 30 minute cadence, the distribution in time used in our analysis was not even, varying from 0.23 to 64 minute time steps; the average time step was 11.4 minutes. These intermediate signal data are created by interpolating in time, which enables sufficient resolution of the ICME structure near Line 4 without increasing computation intensity for the remaining 75 hours of the simulation as the ICME propagates outward across the other Lines 1-3. Note that in calculating the FF, the TEC samples had to be interpolated in time to enable a consistent rate of 100 seconds. Variables from each voxel in the AWSoM simulation were utilized for different aspects of this work; they were the magnetic field vector, density, Alfven wave energy, temperature, and the velocity vector.

The magnetic field and density variables were used to calculate the FR and TEC on each Line 1-4. Light-time was accounted for in these calculations with the values in the voxel traversed being updated with an interpolation with its value in the next time step. Line 1's light-time was the shortest at 14.434 minutes, Lines 2 and 3 were 15.758 minutes, and Line 4 was the longest at 16.504 minutes. For the FF calculations, the values were too smooth, so the Alfven wave energy was used to roughen up the resulting change in TEC as described in section \ref{sec:anothersubsection}. The temperature and velocity vector were used only at the simulated in-situ spacecraft observing for magnetic flux rope analysis.

\subsection{Synthetic in-situ Data}\label{sec:synthdata}
\begin{figure}
    \begin{tabular}{cc}
    \includegraphics[width=3.5in]{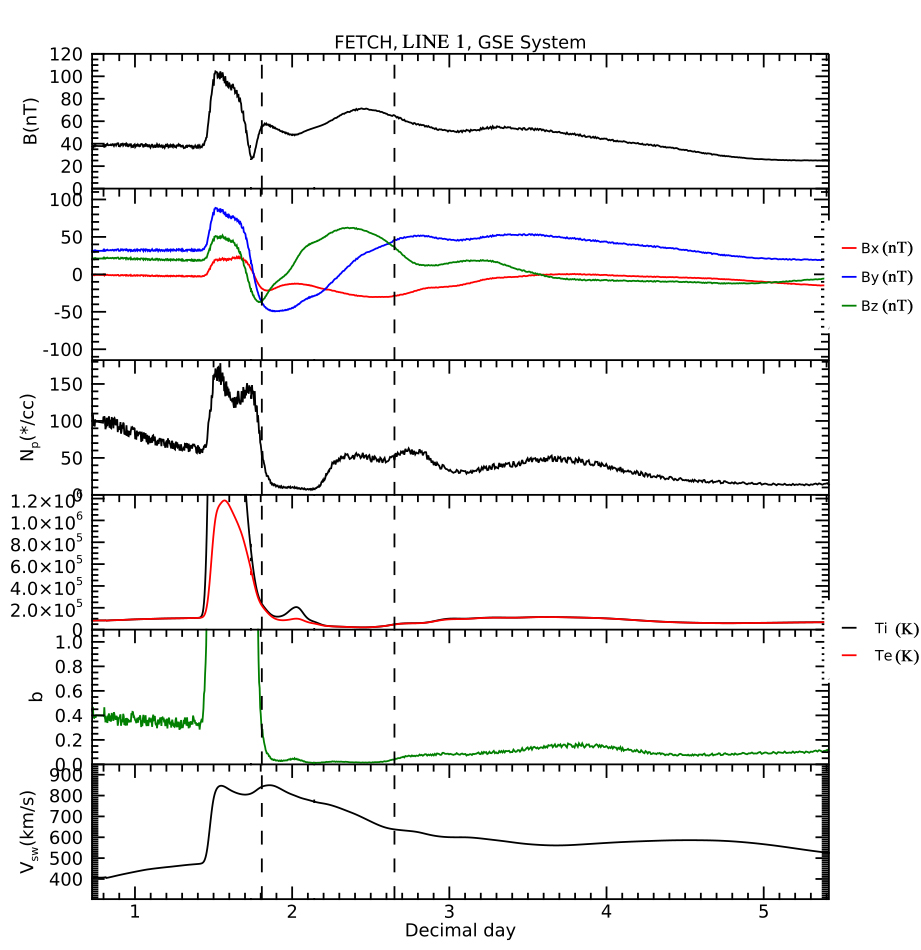}             
   & 
 \includegraphics[width=3.5in]{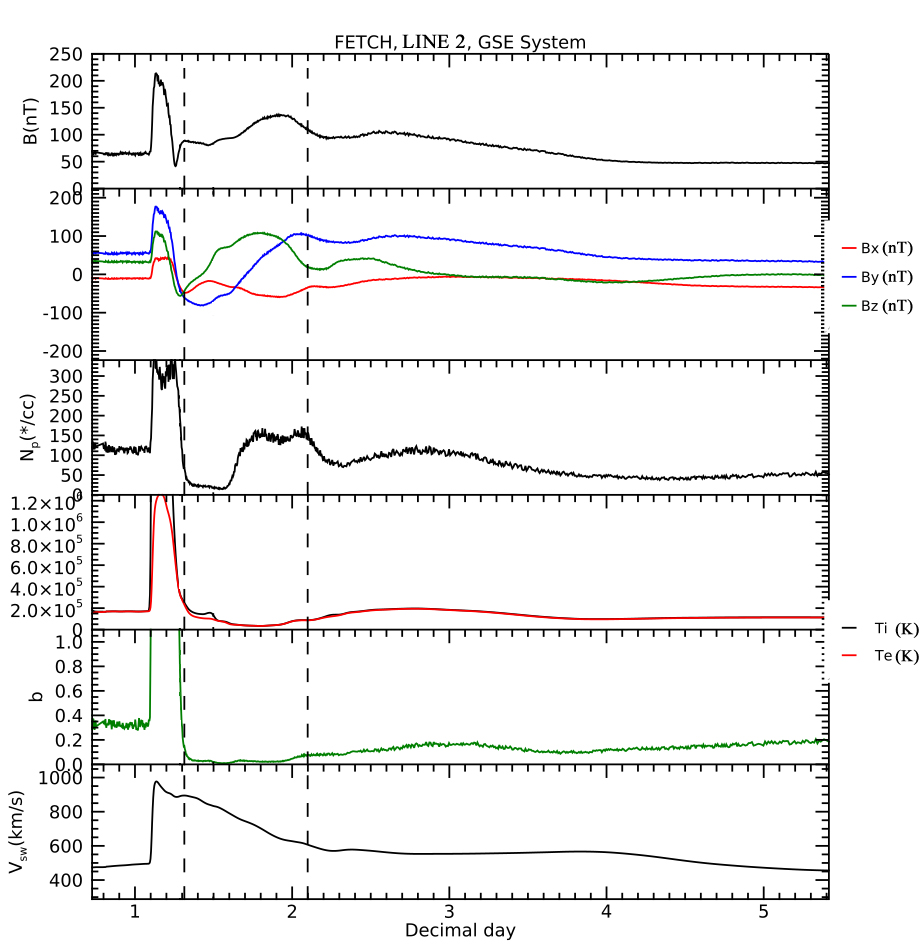}          
  \\
    \includegraphics[width=3.5in]{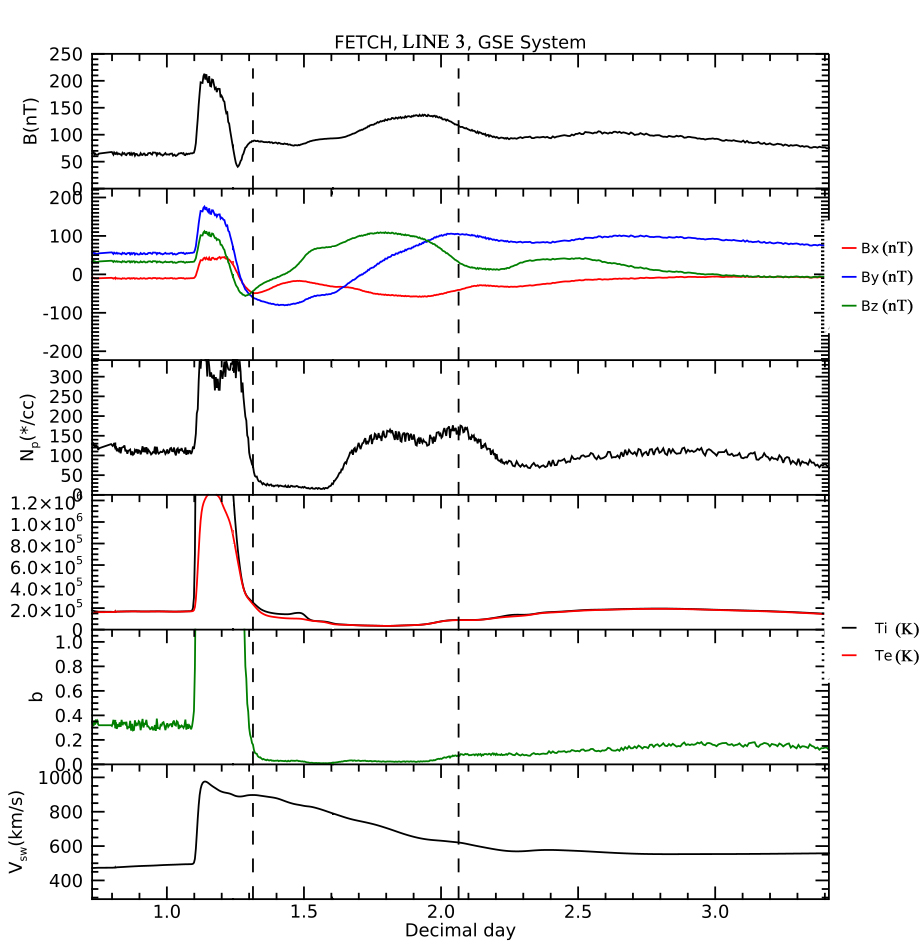}   
    & 
 \includegraphics[width=3.5in]{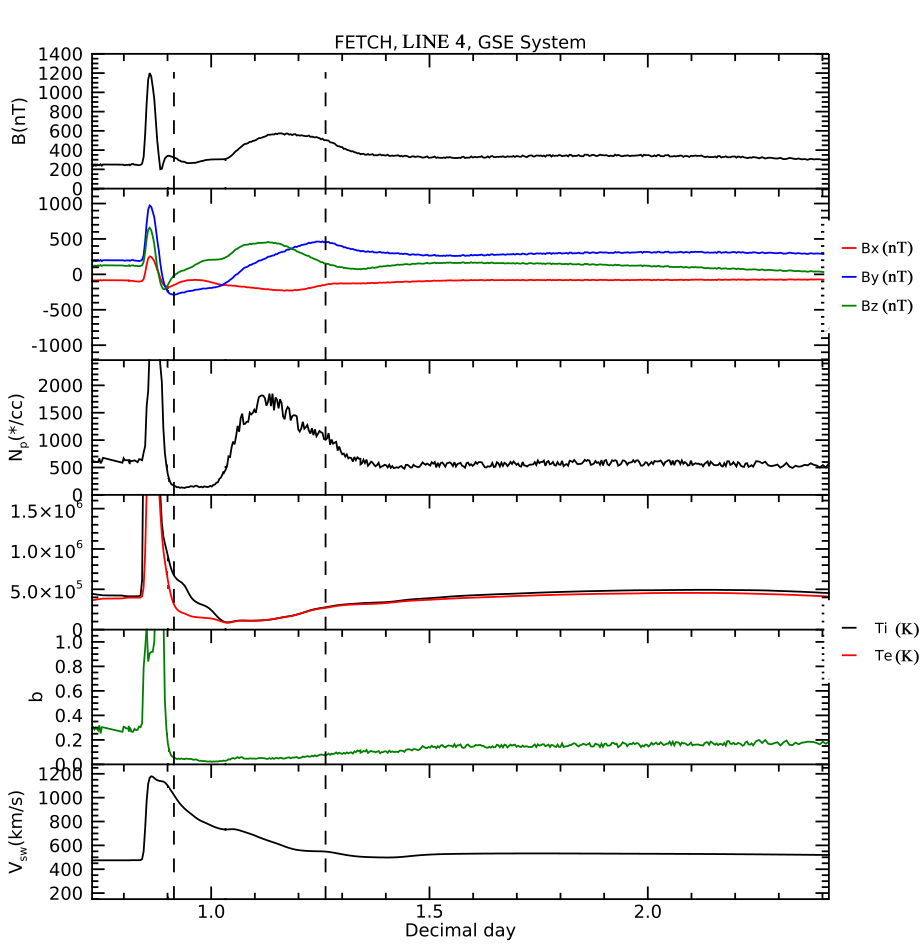}
    \end{tabular}{cc}
    \caption{{\em In-situ} time series at the LOS closest approach on each FETCH Line, labeled 1-4. Magnetic field (GSE), density, temperature, plasma beta, and velocity are shown with time. The dashed lines show the leading/trailing edge of the magnetic flux rope based on plasma properties and magnetic field.}
    \label{fig:plasmalines1-4}
\end{figure}

\begin{figure}
    \centering
    \includegraphics[width=6in]{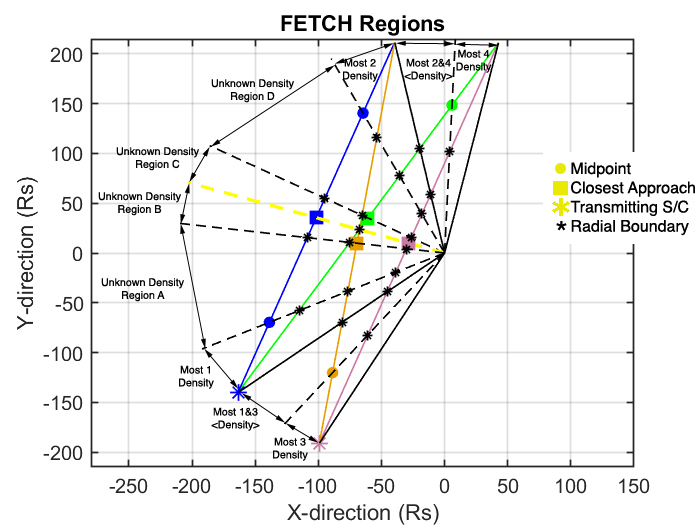}
    \caption{FETCH analysis layout. The trajectory of the ICME (yellow line) and the offset point locations (colored squares) for the lines-of-sight are shown. Line 1 (solid blue), Line 2 (solid green), Line 3 (solid orange), and Line 4 (solid purple) are shown. The colored circles show on the line where there is equal sum of offset distances on either side between the MOST spacecraft and point of closest approach. MOST 1 and 3 are transmitting with MOST 2 and 4 receiving in this paper's analysis. The black * in the figure show the boundary crossings of the different density regions on each line-of-sight. Note that `density' means weighted by $r^{-2}$ from the Sun. Appendix \ref{app:tomo} shows the details for how this is used in the tomographic analysis.}
    \label{fig:linescomp}
\end{figure}

With the ICME's crossing each line of sight, {\em in-situ} data was extracted from the AWSoM 2005 model at the point of closest approach (shown in colored squares in Figure \ref{fig:linescomp}). These points in the simulation were selected to investigate what a spacecraft would observe if positioned in those locations. The simulated spacecraft time series observations consisted of the magnetic field magnitude, components, plasma density, proton/electron temperatures, and velocity (see Figure \ref{fig:plasmalines1-4}). The magnetic flux-rope (MFR) portion is delimited by two dashed lines in the panels of Figure \ref{fig:plasmalines1-4}. Based on visual inspection of a combination of plasma quantities over the entire duration of the ICME passage, each Leading-Edge (LE) and Trailing-Edge (TE) are determined. The criteria combine the magnetic field direction rotation, the crossing of the Bz component from south to north, lower temperature and plasma beta (b), and the average solar wind velocity ($V_{sw}$) (see \citealp{2006SoPh..239..393J} or \citealp{2022ApJ...930...88N} for more details). The magnetic field strength enhancement indicates a bit of compression at the back of the structure.  From Line 4 to Line 1, there is an increase in the magnetic field strength after the flux rope. Note that the small change in axial position between the location at the closest approach on Line 2 and Line 3 produce different in-situ structures.

There is a significant issue to be aware of with regard to timing the ICMEs leading/trailing edges by comparing {\em in-situ} data to FR/TEC integrated data. The LOS integrated data will detect the coherent structure when enough of its structure fills the LOS. Regardless of the ambiguity in what direction the ICME travels, its time of passage is detected as it crosses the LOS. In the case of {\em in-situ} data, the observation is sensitive to the local structure enveloping the observing spacecraft. If the path of the structure is not centered on the spacecraft, then its leading/trailing edge times will reflect this.

\subsection{FR Analysis of the Simulated Event}\label{sec:anothersubsection}
In practice the FETCH signals would be travelling both ways between spacecraft to measure the plasma between them (over the $<$16 minute light time) ; however, we are not resolving/exploiting this tomographic capability as described in \cite{FUNG2022101}. To keep it simple and applicable to all previous and current observations, the signals are only being transmitted by MOST 1\&3, received by MOST 2\&4.

Each LOS provided a unique measure of the ICME. Recall that in this simulation, the light-times are accounted for. Line 1's light-time was the shortest at 14.434 minutes, Lines 2 and 3 were 15.758 minutes, and Line 4 was the longest at 16.504 minutes. The LOS between MOST 1\& 4 (Line 2) and 2\& 3 (Line 3) provided different measures of the ICME structure with its path offset from the Earth-Sun line by 78 and 102 degrees (as opposed to 90 degrees on Lines 1 \& 4). Line 2 integrated less of the sheath and flux rope core together in its measurement than Line 3. Lines 1 \& 4 measured the evolution of the ICME structures with the same view angle. The locations of MOST 3 and 4 were set to equidistant from the Earth with an impact parameter for Line 4 of 30 solar radii from the Sun (see Figure \ref{fig:MOST} LOS Line 4). With respect to the coordinate system for the following analysis, it is heliographic inertial (HGI).

Noise was added to the simulation model to roughen up the result; the simulated FFs were too smooth. Alfv\'en wave energy from the AWSoM model ($E_w$ expressed in $erg/cc$) was used to calculate the fluctuations in magnetic field strength with the equation $\sqrt{8\pi E_w}$; this was approximately 10\%, so a uniform distribution random number within this value was added to the simulation value. The density was similarly estimated to vary by around 10\%, from Parker Solar Probe {\em in-situ} proton measurements (see the analysis in Appendix \ref{app:instvel}). A more realistic variability is not possible with these tools at present; combining instability models with an MHD model is a major undertaking. For instance this variability was found to be incapable of reproducing actual observational variability in TEC and FR.

\subsubsection{Examining Parallel Magnetic Field Calculation}\label{sec:ave_B}
\begin{figure}
    \centering
    \includegraphics[width=6.5in]{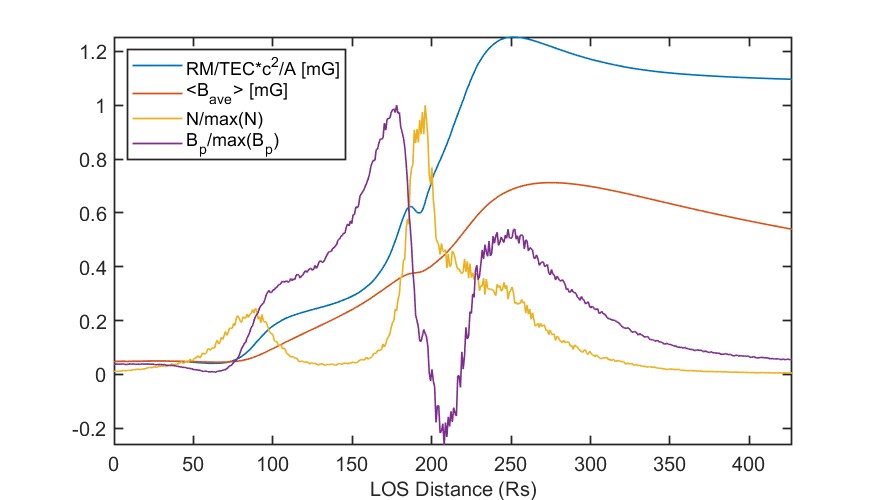}
    \caption{Comparison of the average parallel magnetic field ($B_p$) as calculated from FR \& TEC (red dashed line) versus the model's solution (blue line) over a single propagation path initialized at one point in time on Line 4 (shown in Figure \ref{fig:linescomp}). Recall that FR and TEC are integrated quantities, so the $B_p$ calculation and TEC are cumulative along Line 4. The density (brown line) and parallel magnetic field (light blue line) along the LOS are plotted, normalized to 1. When the density regionally increases, the differences between the two averages increases.}
    \label{fig:bpcomp}
\end{figure}

A very common approach in radio remote sensing is to estimate the average magnetic field using TEC data. This simulation is the first time that this approximation has been examined. When we compared the simple technique for calculating the parallel magnetic field $<B_{\mathrm{\parallel}}> \approx FR / TEC \times f_0^2 / A$ to the actual average parallel magnetic field in the simultion, Panel R3 in Figures \ref{fig:Line1} through \ref{fig:Line4} show that these values were off by roughly a factor of 2. Closer analysis revealed that the discrepancies developed in regions of increased density. Figure \ref{fig:bpcomp} shows how the difference in $<B_{\mathrm{\parallel}}>$ developed over the LOS for a single point in time. The concentration of density aligned with a reversal of magnetic field weighting against the ultimate FR, while enhancing the TEC. This indicates that a tomographic analysis of the TEC could improve the measurement of the average magnetic field in the LOS, an endeavor for a separate paper. 

Recall that we don't have any remote sensing radio data to compare with this simulation. So while a thorough analysis would be important for fusing data and modeling, it's not appropriate in this context which is entirely based on a simulation. This is an inspection of what could be possible with a FETCH-like instrument.

\begin{table}
\begin{tabular}{c|c|c|c|c|c|c}
  \begin{minipage}[c]{1cm}Offset\end{minipage} & \begin{minipage}[c]{1.7cm}RM Line\\4$\rightarrow$3$\rightarrow$1\\ \end{minipage} & \begin{minipage}[c]{1.7cm}TEC Line \\4$\rightarrow$3$\rightarrow$1\\ \end{minipage} & \begin{minipage}[c]{2.1cm}{\em in-situ} structure timing \\4$\rightarrow$3$\rightarrow$1\\ \end{minipage} & \begin{minipage}[c]{1.7cm}RM Line\\4$\rightarrow$2$\rightarrow$1\\ \end{minipage} & \begin{minipage}[c]{1.7cm}TEC Line \\4$\rightarrow$2$\rightarrow$1\\ \end{minipage} & \begin{minipage}[c]{2.1cm}{\em in-situ} structure timing \\4$\rightarrow$2$\rightarrow$1\\ \end{minipage} \\ \hline \hline 
  \multicolumn{7}{c}{Timed sheath speed (km/sec).}\\
    50 & 1100 & 1027 & 984 & 1062 & 916 & 1075 \\
    89 & 648 & 648 & 885 & 663 & 706 & 818 \\ \hline
  \multicolumn{7}{c}{Sheath speed fraction of averaged in-situ speed.}\\
    50 & .98 & .92 & .88 & 0.99 & .85 & 1.0 \\
    89 & .68 & .68 & .93 & .73 & .78 & .91\\ \hline \hline
  \multicolumn{7}{c}{Timed interior flux rope speed (km/sec).}\\
    50 & 566 & 522 & 540 & 531 & 566 & 525 \\
    89 & 518 & 416 & 564 & 554 & 390 & 583 \\ \hline
  \multicolumn{7}{c}{Interior flux rope speed fraction of averaged in-situ speed.}\\
    50 & .80 & .74 & .77 & .75 & .80 & .75 \\
    89 & .70 & .56 & .77 & .75 & .53 & .79 \\ \hline \hline 
  \multicolumn{7}{c}{Timed trailing edge speed (km/sec).}\\
        50 & 421 & 425 & 404 & 378 & 358 & 388 \\
    89 & 414 & no feature & 521 & 470 & no feature & 554 \\ \hline
  \multicolumn{7}{c}{Trailing edge speed fraction of averaged in-situ speed.}\\
    50 & .71 & .72 & .68 & .63 & .60 & .65 \\
    89 & .65 &     & .82 & .74 &     & .87
\end{tabular}
\caption{In each section: Top two rows show Timing speeds in (km/sec) and the bottom two rows show their fraction of the {\em in-situ} speed. The offset values are averages between the offset points (in solar radii) $<(30,70)>=50$ and $<(70,108)>=89$. The table {\em in-situ}-column values are based on averaging arrival times of the MFR structure of interest at the points of closest approach. Note that TIMING speeds in the simulation at the points of closest approach are different from the AVERAGE {\em in-situ} ICME-path-of-travel velocity vectors. For the flux rope interior, the table {\em in-situ} values are based on averaging the LE and TE of the MFR. For the TE, TEC and RM times are based on the enhancement from the prominence material trailing the MFR.}
\label{tab:veltables}
\end{table}

\begin{figure}
    \centering
    \includegraphics[width=6in]{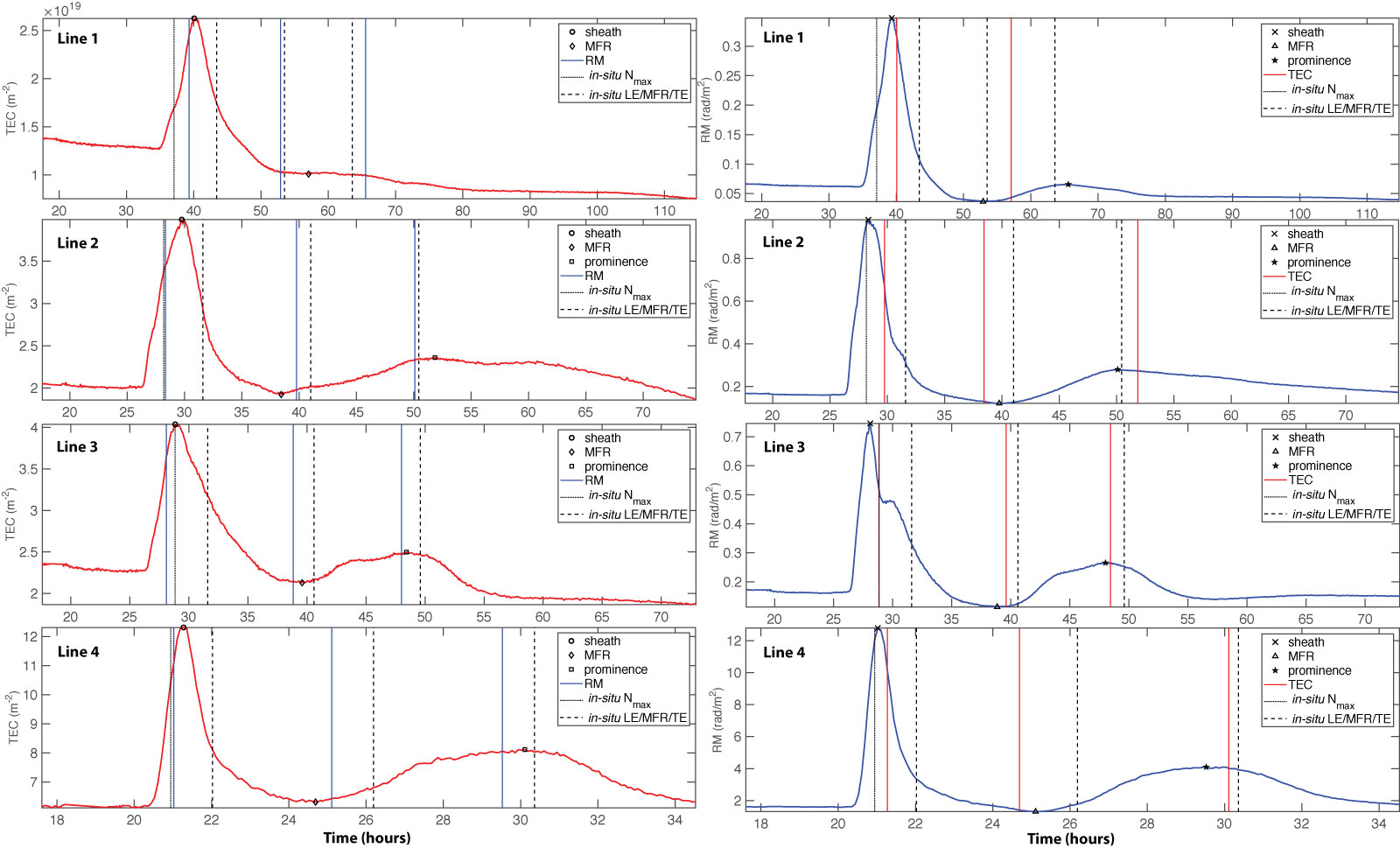}
    \caption{ICME passage across the Lines 1-4. The symbols denote time location of the TEC/RM maximum, minimum, and prominence bump on the left/right respectively. The blue/red lines overlay these on the left/right for comparison. The black lines give the {\em in-situ} maximum density, and the MFR leading edge, center, and trailing edge. The ordinate label identifies the plot as TEC or RM. The black symbols on the red TEC plots are the locations for the red vertical lines on the RM plots and vice versa. Note that Line 1 TEC has no feature for the prominence.}
    \label{fig:speedcombo}
\end{figure}

\subsubsection{Examining Velocity/Speed Observations}\label{sec:speed_stuff}

Speed/velocity measurements are obtained in a variety of ways with radio and spacecraft. As mentioned previously, the AWSoM model calculates plasma bulk velocity from which the LOS average velocity is extracted for analysis. Examples of the average velocity components are shown in Figures \ref{fig:plasmalines1-4}. For the analysis of what the FETCH spacecraft would observe, the only direction that is retained is perpendicular to the LOS, so we report these measures as speed. It is possible to track a feature, such as the sheath of the ICME, from one LOS to another, thereby producing `timed' measurements of speed. The aim of this section is to compare the average velocities with various timed measurements.

The next step in the ICME analysis is to measure its speed. This enables comparing different approaches against the simulation and estimating the size of the MFR later. Note that estimating velocity requires a supplementary calculation for the propagation direction off the Earth-Sun line, which is beyond the scope of this paper. The 2005 ICME we are simulating was a direct impact on the Earth. There are indications that this offset parameter matters in a comparitive sense, namely the comparison of Lines 2 and 3 which are at small offset angles to the perpendicular (90 degrees) of the Earth-Sun line, 78 and 102 degrees. The minimum difference in speed between Lines 4$\rightarrow$2$\rightarrow$1 and 4$\rightarrow$3$\rightarrow$1 was 15 km/sec; the maximum was 111 km/sec. The lack of consistency in the speed variations between Lines 4$\rightarrow$2$\rightarrow$1 and 4$\rightarrow$3$\rightarrow$1 indicates that the asymmetric detection of the ICME structure is important for this kind of analysis. This result supports the tomographic approach taken by \cite{Jackson:1998}.

As shown in Figure \ref{fig:speedcombo}, the ICME is comprised of unique structural characteristics which can be identified each time it crosses a different LOS, specifically the sheath, the interior of the flux rope, and the trailing edge of the flux rope. In order to inspect measuring speed by timing these structures crossing each LOS, we needed to characterize how the speed measurement of the ICME was different from the {\em in-situ} speed at the points of closest approach. One detail to keep in mind when comparing {\em in-situ} data to columnar integrated data is that the latter requires enough of it to be crossing the line-of-sight, and in the case of the magnetic field to be coherent over a large extent for a structure to become apparent.

While the simulation provided {\em in-situ} average velocities that can be compared, the information that we needed was to perform a similar timing exercise for when notable portions of the structure crossed each LOS. We needed to identify the sheath, interior of the flux rope, and the trailing edge of the flux rope {\em in-situ} and time how long it took to appear {\em in-situ} at the next spacecraft placed in the simulation's point-of-closest approach for Lines 1 to 4. This comprises columns 4 and 7 in Table \ref{tab:veltables}.

Identifying the regions was not consistent between the data sets. First of all definitions of each region depended on the time series. The sheath was identified as the maximum TEC or RM. The flux rope was identified by the minimum in the TEC or RM. Finally, the trailing edge {\em FOR THIS SPEED CALCULATION ONLY} was identified as the `bump' caused by the presence of the trailing prominence material. The TEC time series for Line 1 was not sensitive to this structure. Consequently, the times for the sheath, flux rope, and trailing edge in TEC were not the same for the RM. These were also not the same as {\em in situ}. 

In order to discern between the LOS integrated measures of TEC/RM versus {\em in situ}, we had to time the structures in the TEC/RM with {\em in situ} variants. We timed the flow of the ICME's maximum density (from the sheath) in the {\em in-situ} measurements for the sheath values. For the {\em in-situ} timing MFR timing, the flux rope point was set to the midpoint between the leading and trailing edges of the flux rope. As discussed, these edges were selected by analysis of simulated {\em in-situ} plasma data. Finally, the trailing edge of the {\em in-situ} structure was selected for timing. The leading edge of the {\em in-situ} flux rope was not used for this analysis.

The {\em in-situ} average velocities in these ICME structures were also compared to the calculated speeds. The result is compared to the other calculations on the rows with the fraction calculations in Table \ref{tab:veltables}. To make the comparison, we averaged the component of the average velocity values perpendicular to Lines 1 to 4. At the time a structure of interest was observed in the {\em in-situ} simulated data, its time was noted as well as its perpendicular average speed. Then the average velocities were averaged, $<V_{Line1},V_{Line2}>$, $<V_{Line1},V_{Line3}>$, etc., giving a different speed. These `averaged in-situ speeds' are the denominator of the fraction values of the table.

As shown in Table \ref{tab:veltables}, the velocities of the AWSoM model ICME regions (sheath, flux rope, and trailing prominence/edge) were between 0.53 and 1.0 of the average velocity/speed parallel to the ICME's path. Based on the RM measures, we find that the ICME is expanding around 350 km/sec initially and around 100 km/sec later. In comparison, the TEC expansion is around 300 km/sec initially. Finally, the {\em in-situ} expansion initially is around 325 km/sec initially and 150 km/sec later; however, there's more variance to the {\em in-situ} timing measurements than to the columnar integrated timings. This is a useful illustration of how expansion speeds can be measured and that the measurements may not agree between the different techniques. It is particularly important that the large differences in {\em in-situ} measurements, being in the path of one portion of the asymmetric ICME as opposed to a different part of the structure, are smoothed out with the columnar integrations.

The RM performed as well as the timed {\em in-situ} values for providing the speed of the ASWoM model MFR; the TEC velocities suffered at 89 Rs when the minimum was no longer distinct as the structure evolved. How relevant this is to actual observations as opposed to this model simulation is unknown.

The {\em in-situ} plasma data determined trailing edge fell near the increase in TEC and RM in the prominence material that occurs following the passage of the dark core. While this point was easy to detect in the RM, at 108 Rs on Line 1 it was not distinguishable from the background TEC. The consequent differences in timing speed between the RM and the {\em in-situ} values indicates that the trailing edge speed timing could be off by 2-17\% relative to the {\em in-situ} timing measures. This is a similar \% for all of the structure speeds for comparing RM and {\em in-situ} timing.

Other details in the model include the sheath region between Lines 4 and 2/3 (50 solar radii average impact parameter) was moving close to the average wind speed; however, it slows between Lines 2/3 and 1 (89 solar radii average impact parameter). Whether or not this is accurate as discussed previously regarding the ICMEs speed in the model, the simulated data from a FETCH spacecraft layout enabled recovering it.

The difference between {\em in-situ} average speed and timing from {\em in-situ} MFR structure suggests two things. The first would be to review multi-spacecraft observations of ICME crossings for differences in average speeds versus timing speeds. The second is that it would be a benefit to be able to obtain from the radio frequency signal more real-time measures of speed averaged across the line-of-sight to compare to timing speeds.
We investigate this possibility in Appendix \ref{app:instvel}. Developing the capability for a FETCH-like instrument to obtain LOS velocities would require utilizing a known technique in interplanetary scintillation. Empirical observations of intensity fluctuations from density changes show that they have a turbulence scaling property that can be utilized for velocity estimates. This technique as described in the Appendix, should be investigated for potentially obtaining average speed measurements from the radio frequency signal. We were unable to simulate this feature to radio observing, because the resolution required (of order 100km) is infeasible.

These differences in calculated velocity provide an important measure for determining the kinetic energy in an ICME. Teasing out any differences between the timing speeds of ICME structures, any differences between the TEC, FR, and {\em in-situ measurements}, and obtained average speed observations enable understanding more about the influence of structural asymmetry in the measurements.

\begin{figure}
    \centering
    \includegraphics[width=0.4\textwidth]{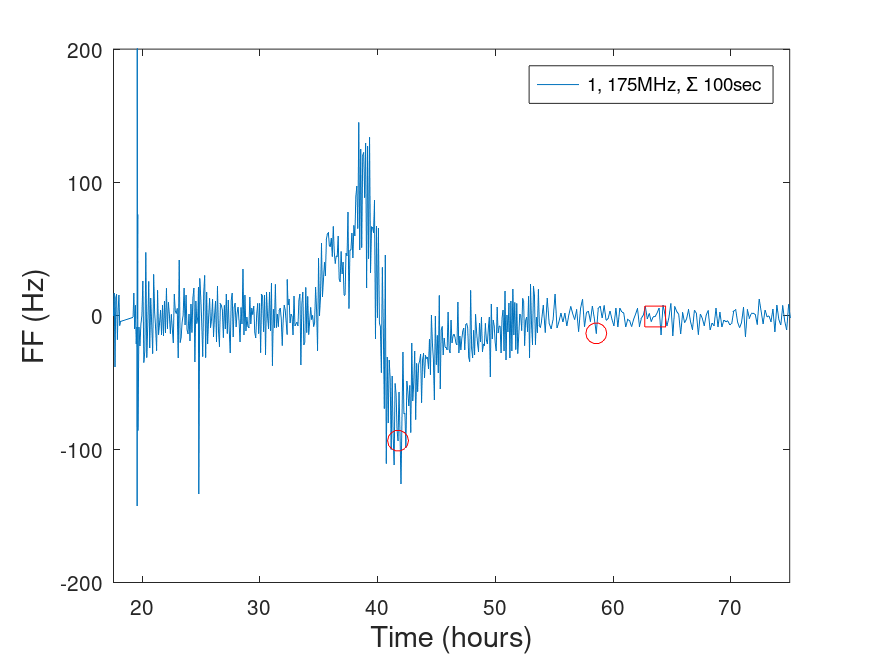}
    \includegraphics[width=0.4\textwidth]{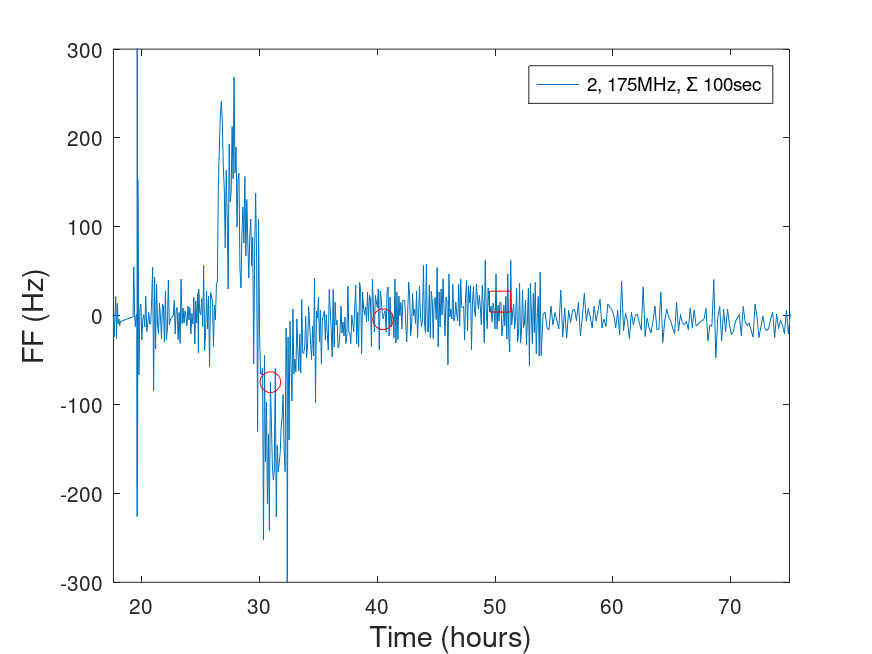}
    \includegraphics[width=0.4\textwidth]{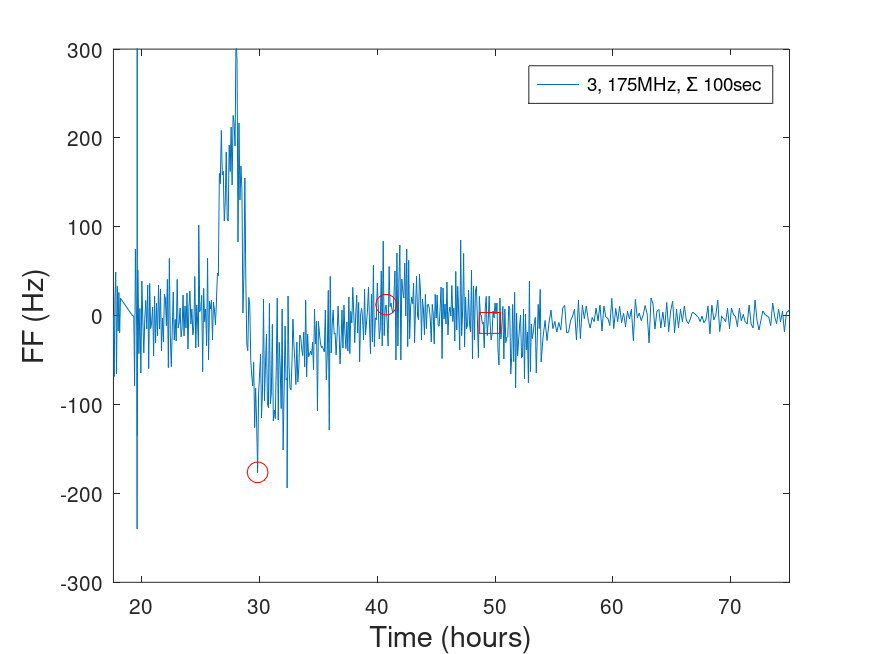}
    \includegraphics[width=0.4\textwidth]{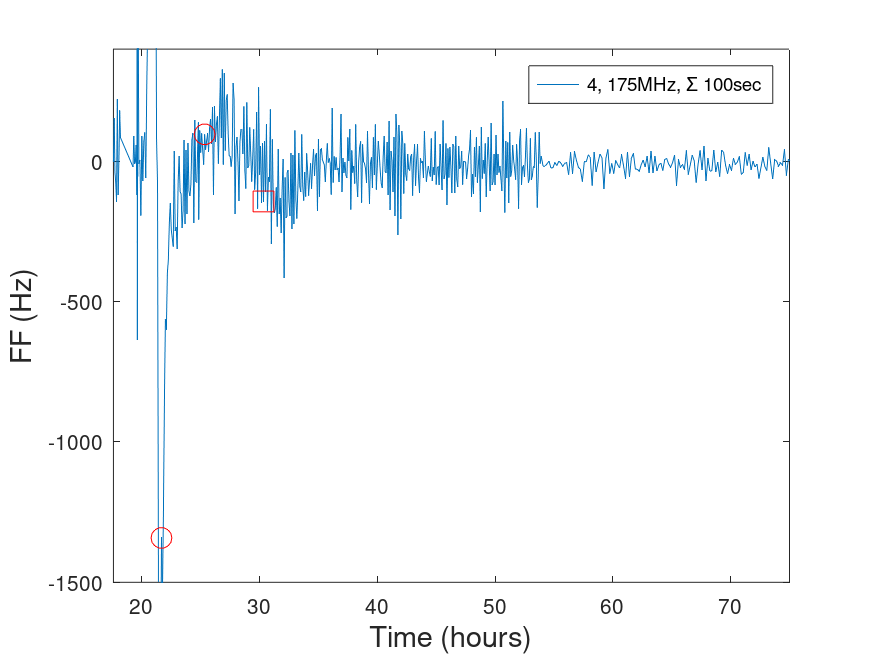}
    \caption{At 165MHz and 100 second integrations, the expected Frequency Fluctuation response. Circles (squares) show the placement of the LE (TE) from regions of minimum (maximum) FF where the TEC is decreasing (increasing) at the fastest rate.}
    \label{fig:ffcomp}
\end{figure}

\subsubsection{The ICME Trailing Edge}\label{sec:trailing_edge}
Measuring the time of the ICME crossing requires determining the Leading-Edge (LE) and Trailing-Edge (TE) to the magnetic flux rope. The crossing time is an important parameter for the next paper in which we perform flux rope fitting to the TEC and RM simulated data. The size of the ICME impacts the strength of its magnetic field for producing the observed FR. The longer the integration baseline is to obtain the observed FR, the weaker the magnetic field has to be mathematically. Similarly, placing the TE in the incorrect point on the time series alters the orientation of the fitted flux rope. Obtaining the correct TE is an important endeavor for studying the structure of the MFR.

The Leading-Edge is located in close vicinity to the sheath {\em in-situ}, or in the case of LOS observations, within it on the descending TEC side. It is visible in the expected FF observations. As shown in Table \ref{tab:selected times}, the LE times correspond closely between this technique of FF and the {\em in-situ} analysis of simulated Sun-Earth spacecraft data.

An issue arises with identifying the TE of the flux rope. Using the simulated {\em in-situ} data, we timed the LE and TE locations of the ICMEs magnetic flux rope. Then we evaluated other techniques. We briefly inspected what simulated white-light data would produce and found that in this particular case three TE solutions would be possibilities. We also investigated the utility of FF changes for identifying this part of the ICME structure; it performed well for identifying the LE. While the FF solution was unique, it was also strongly dependent on the AWSoM simulation with regard to the TE as shown in Table \ref{tab:selected times}. Other issues with using FF in this manner to identify the TE are (a) not all CMEs/ICMEs have trailing prominence material and (b) placing the TE at this location was not an initial selection by the authors before we had the MFR analysis.

With the expected available data, RM, TEC, FF, and 9 other instrument suites on MOST, one potential \emph{qualitative} approach is to survey the situation and select the solution that is unique to the conditions of the ICME's propagation. The qualitative approach would develop a checklist using patterns in the empirical simulated observations and real observations when available. While the pattern-recognition approach is common, there is no guarantee it would work. Therefore, we have found an alternate approach to identifying a unique solution utilizing FETCH's layout across the Earth-Sun line: low resolution tomography. 

\begin{figure}
    \centering
    \includegraphics[width=3in]{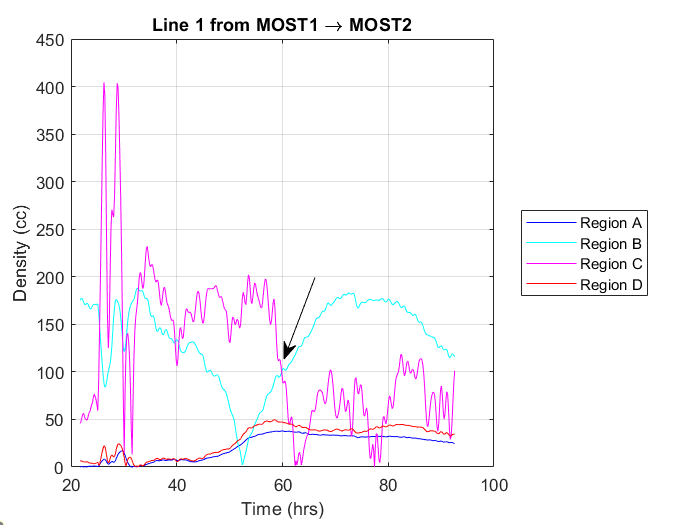}
    \includegraphics[width=4in]{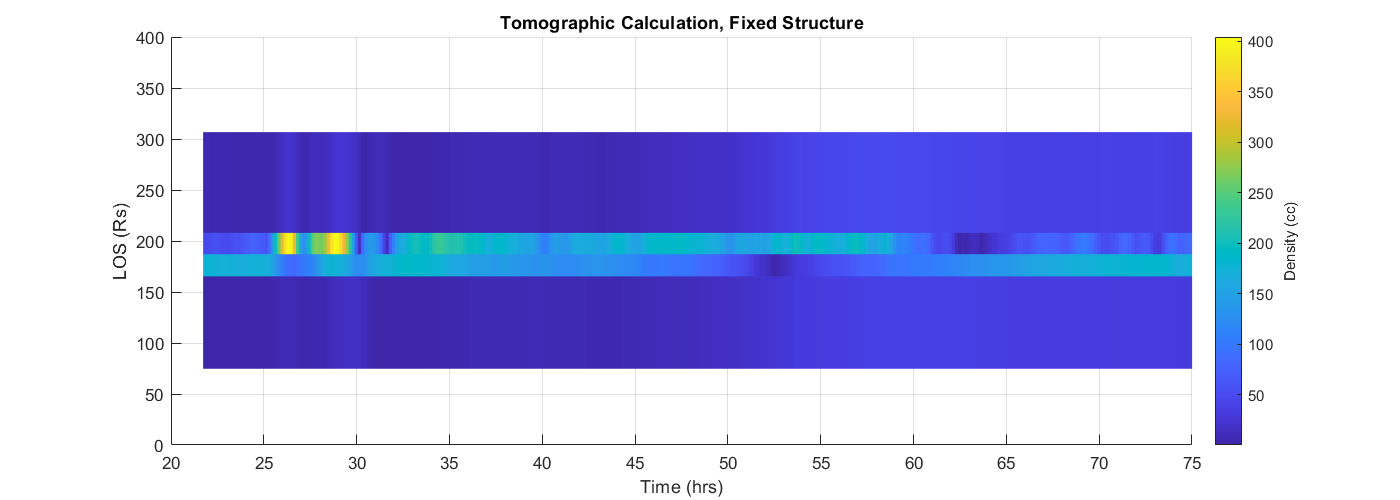}
    \caption{Tomographic analysis via time series mapping. The left-side image shows the values which are plotted spatially on the right-side image. The regions are defined in Figure \ref{fig:linescomp}. The dark core region is centered near 55 hours. The timing shown is in Line 2 and 3 time and requires translation to Lines 1 and 4 (recall the irregular time steps). The black arrow shows where the trailing edge of the flux rope is set based on the peak plasma densities calculated with this technique in Region C after the passage of the ICME. Appendix \ref{app:tomo} provides the details behind this analysis.} 
    \label{fig:tomogL2}
\end{figure}

\begin{figure}
    \centering
    \includegraphics[width=1.0\textwidth]{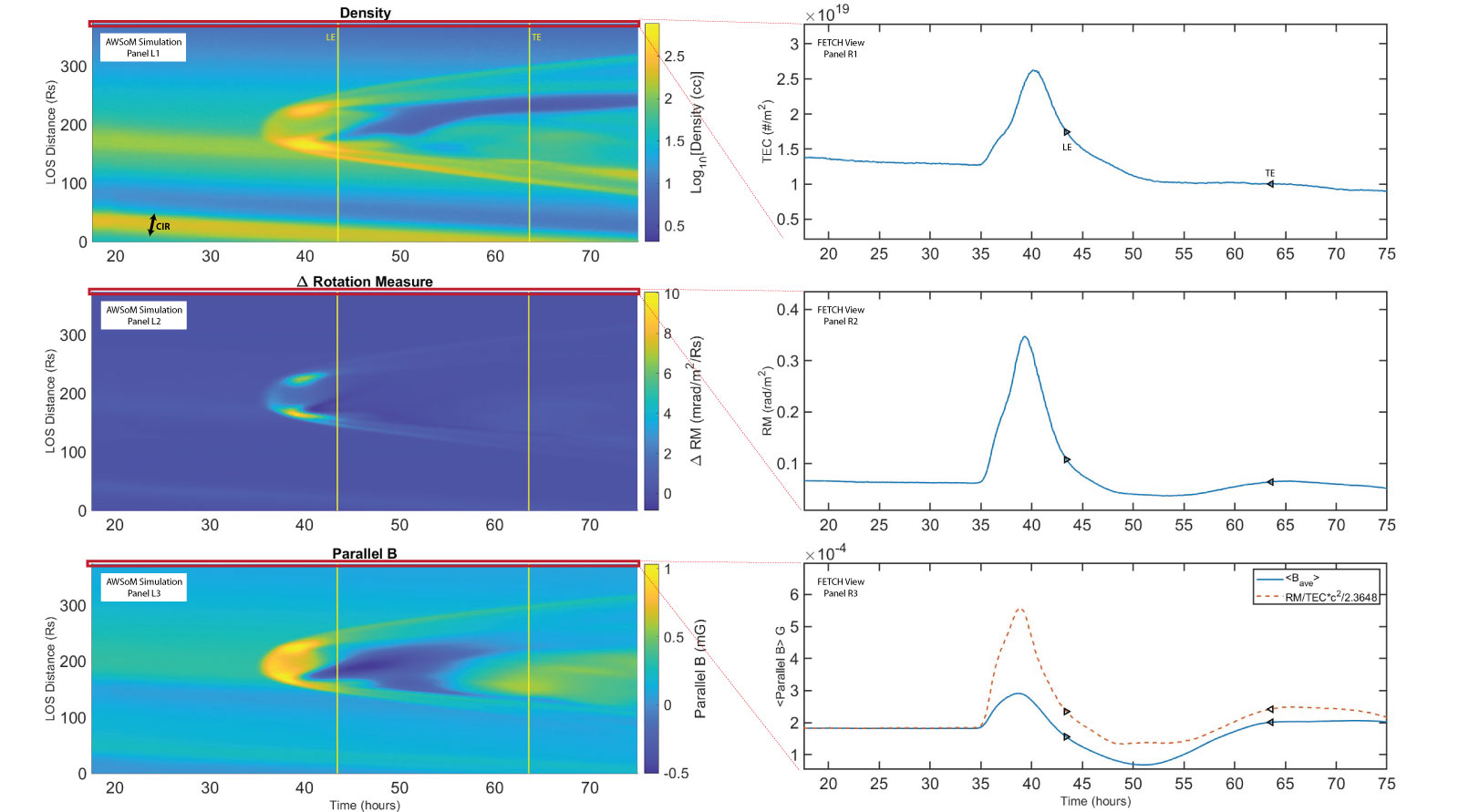}
    \caption{Simulated ICME structure extracted along Line 1 Forward (MOST 1$\rightarrow$2) shown in a time series. Shown from top to bottom respectively are the spatial distribution with time of the Density (L1, Corotating stream Interaction Region ``CIR" is labeled), the $\Delta RM$ (L2), and the Parallel Magnetic Field to the LOS (L3). The ordinate axis shows the distance along the LOS in solar radii. The plots to the right show the observations FETCH would obtain of the TEC (R1), the RM observed at MOST 2 (R2), and the average magnetic field (both modeled solid-blue and calculated dashed-red, Panel R3). LE and TE show the Leading and Trailing Edges of the ICME from simulated {\em in-situ} analysis.}
    \label{fig:Line1}
\end{figure}

\begin{figure}
    \centering
    \includegraphics[width=1.0\textwidth]{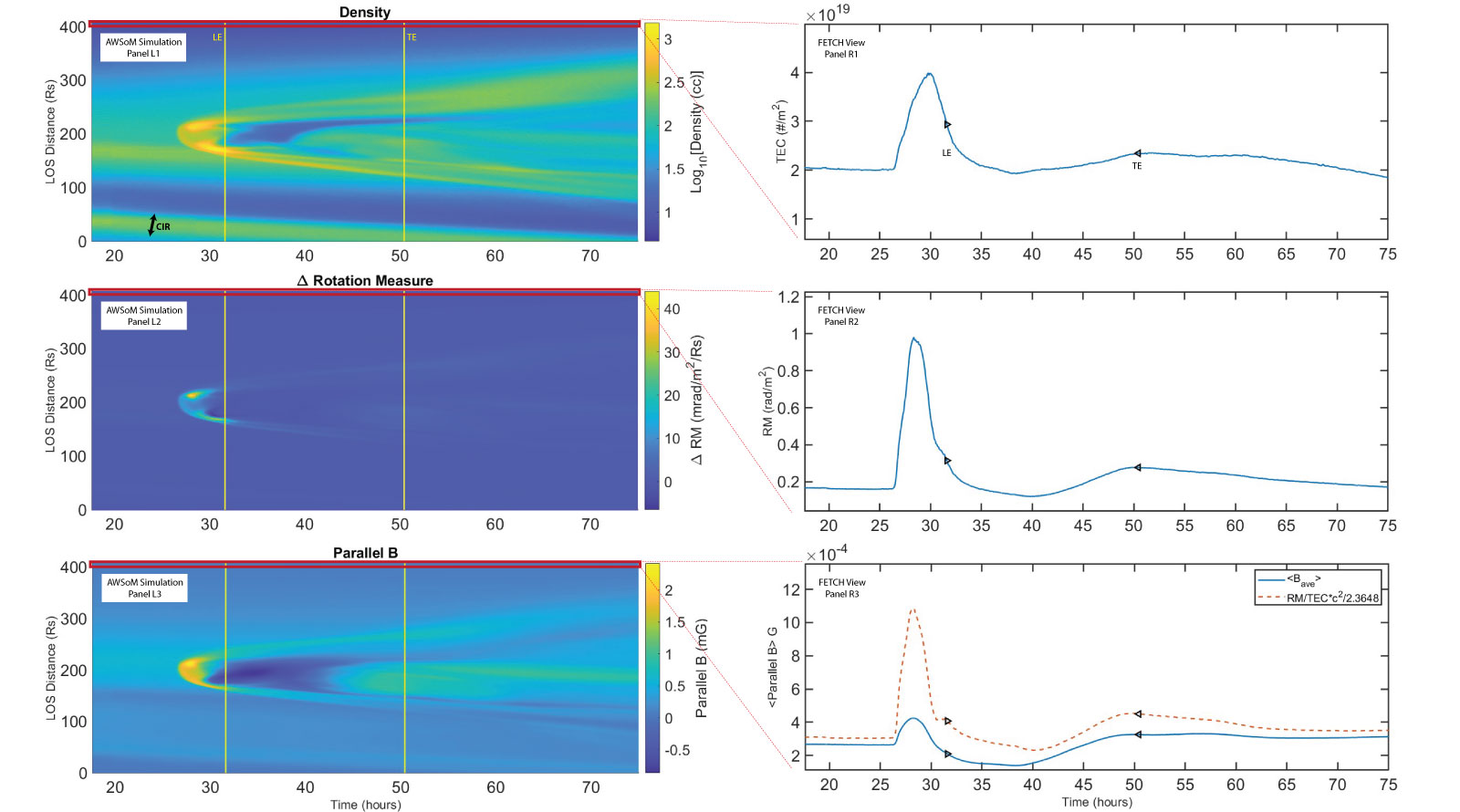}
    \caption{Simulated ICME structure extracted along Line 2 Forward (MOST 1$\rightarrow$4) shown in a time series. Shown from top to bottom respectively are the spatial distribution with time of the Density (L1, Corotating stream Interaction Region ``CIR" is labeled), the $\Delta RM$ (L2), and the Parallel Magnetic Field to the LOS (L3). The ordinate axis shows the distance along the LOS in solar radii. The plots to the right show the observations FETCH would obtain of the TEC (R1), the RM observed at MOST 4 (R2), and the average magnetic field (both modeled solid-blue and calculated dashed-red, Panel R3). LE and TE show the Leading and Trailing Edges of the ICME from simulated {\em in-situ} analysis.}
    \label{fig:Line2}
\end{figure}

\begin{figure}
    \centering
    \includegraphics[width=1.0\textwidth]{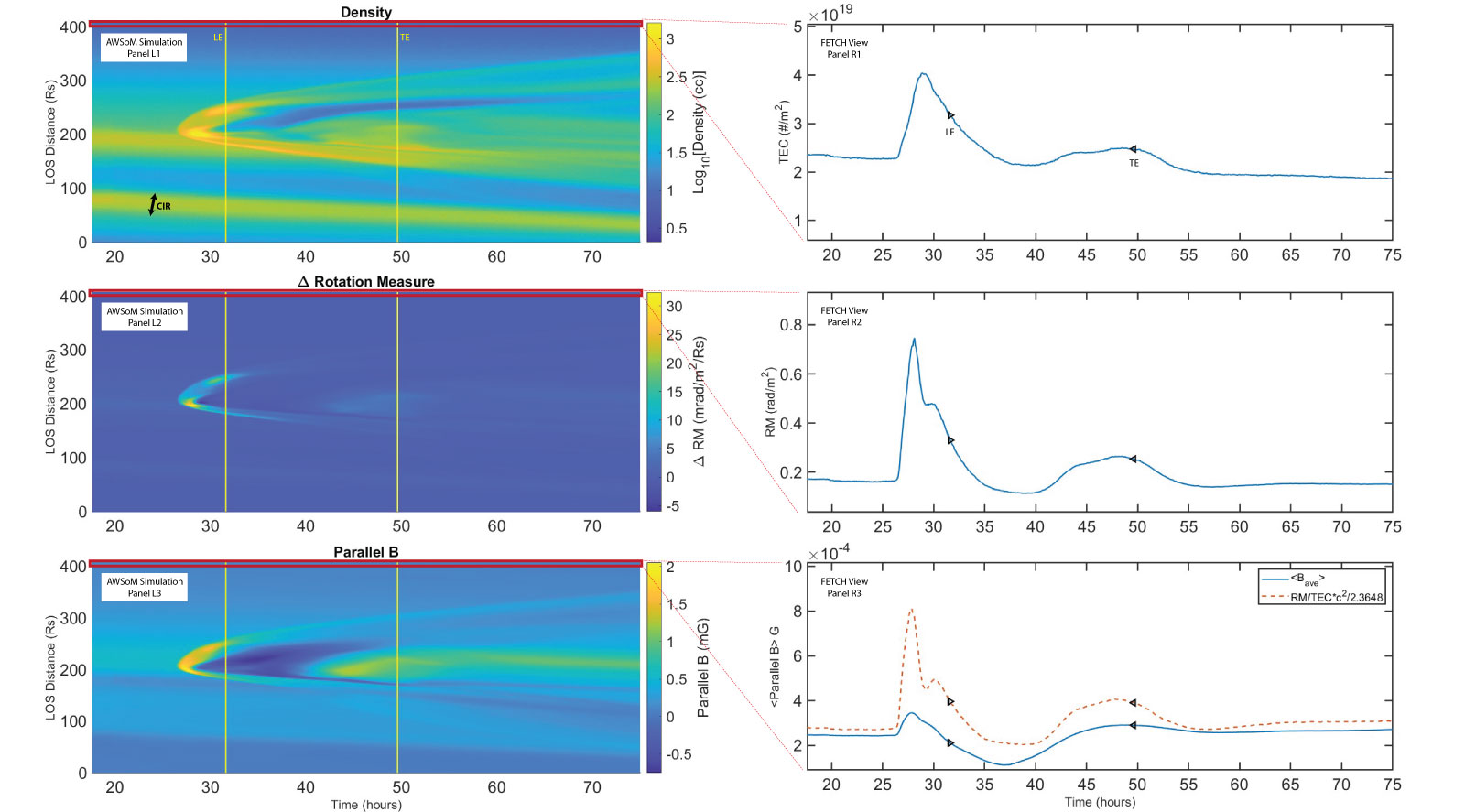}
    \caption{Simulated ICME structure extracted along Line 3 Forward (MOST 3$\rightarrow$2) shown in a time series. Shown from top to bottom respectively are the spatial distribution with time of the Density (L1, Corotating stream Interaction Region ``CIR" is labeled), the $\Delta RM$ (L2), and the Parallel Magnetic Field to the LOS (L3). The ordinate axis shows the distance along the LOS in solar radii. The plots to the right show the observations FETCH would obtain of the TEC (R1), the RM observed at MOST 2 (R2), and the average magnetic field (both modeled solid-blue and calculated dashed-red, Panel R3). LE and TE show the Leading and Trailing Edges of the ICME from simulated {\em in-situ} analysis.}
    \label{fig:Line3}
\end{figure}

\begin{figure}
    \centering
    \includegraphics[width=1.0\textwidth]{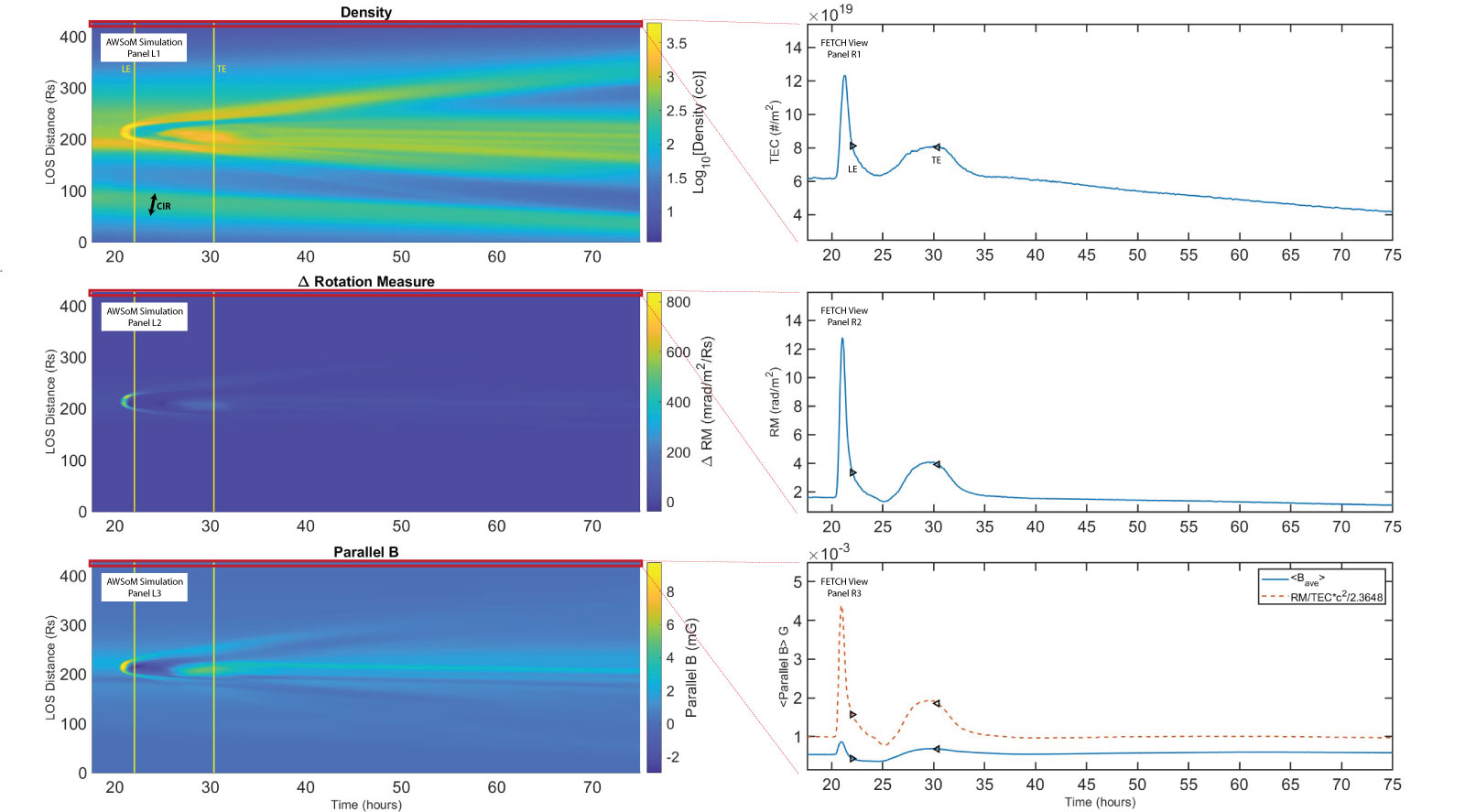}
    \caption{Simulated ICME structure extracted along Line 4 Forward (MOST 3$\rightarrow$4) shown in a time series. Shown from top to bottom respectively are the spatial distribution with time of the Density (L1, Corotating stream Interaction Region ``CIR" is labeled), the $\Delta RM$ (L2), and the Parallel Magnetic Field to the LOS (L3). The ordinate axis shows the distance along the LOS in solar radii. The plots to the right show the observations FETCH would obtain of the TEC (R1), the RM observed at MOST 4 (R2), and the average magnetic field (both modeled solid-blue and calculated dashed-red, Panel R3). LE and TE show the Leading and Trailing Edges of the ICME from simulated {\em in-situ} analysis.}
    \label{fig:Line4}
\end{figure}

In Appendix \ref{app:tomo}, the tomographic analysis is presented. It is a first order solution which yields a distribution in density similar to Panel L1 in Figures \ref{fig:Line1} - \ref{fig:Line4} as shown in Figure \ref{fig:tomogL2}. The unique layout of the FETCH instrument enables the rapid identification of four pie wedges of space with the Sun at the center of the pie. The distribution of density among these wedges can be determined by a straightforward analytical calculation assuming that the structure does not evolve as it crosses the 4 lines of sight. The dark core of the ICME is apparent around 55 hours (in Line 3 time). If we assume that the core is symmetrical in time (uniformly radially expanding), then we obtain the TE values shown in \textcolor{blue}{blue} in Table \ref{tab:selected times}. With respect Lines 1 and 4, these timings perform significantly better than FF. Note that in order to perform the tomographic calculation, the LOS variations with time were stretched and shifted for the solutions shown in Figure \ref{fig:tomogL2}. Table \ref{tab:selected times} shows the de-transformed times. The TEC errors from this approach are shown in Figure  \ref{fig:matrixerr}. At the time of the flux rope crossing, the greatest source of error is Line 4; this is expected since the MFR evolves the most early on in its propagation towards the Earth. Within the MFR, it was 33\% or less in TEC while the resulting density values were within an order of magnitude. Analyzing the impact of these errors in MFR density and size belongs in the MFR-fitting comparison paper.

\begin{table}[]
    \centering
    \begin{tabular}{c|c|c}
        Line & Leading Edge & Trailing Edge \\
        1 & 14/19:19 v \textcolor{red}{14/17:33} v \textcolor{blue}{14/18:09} & 15/15:24 v \textcolor{red}{15/11:07} v \textcolor{blue}{15/15:44} \\
        2 & 14/07:30 v \textcolor{red}{14/07:10} v \textcolor{blue}{14/09:40} & 15/02:20 v \textcolor{red}{15/00:41} v \textcolor{blue}{15/07:45} \\
        3 & 14/07:30 v \textcolor{red}{14/05:49} v \textcolor{blue}{14/11:55} & 15/01:30 v \textcolor{red}{14/18:26} v \textcolor{blue}{15/11:55} \\
        4 & 13/21:56 v \textcolor{red}{13/21:47} v \textcolor{blue}{13/23:21} & 14/06:16 v \textcolor{red}{14/02:46} v \textcolor{blue}{14/06:41}
    \end{tabular}
    \caption{Times from the {\em in-situ} versus \textcolor{red}{FF changes} versus \textcolor{blue}{tomographic analysis}  plasma characteristics. Note that the times provided are specifically in terms of arrival time at the receiving spacecraft. Format day-of-month/hour:minute}
    \label{tab:selected times}
\end{table}

\section{Results}\label{sec:obs_results}
This is the first study to inspect techniques in magnetic field estimating through Faraday rotation analysis in combination with Total Electron Content analysis techniques against a high resolution MHD modeled ICME and heliosphere. The following results were found in this investigation:
\begin{itemize}
    \item Section \ref{sec:ave_B} : The parallel magnetic field approximation of $<B_{\mathrm{\parallel}}> \approx FR / TEC \times f_0^2 / A$ is approximately a factor of 2 off without any further analysis. The error can reach a factor of 5 in regions such as the sheath of the ICME.
    \item Section \ref{sec:speed_stuff} : Speeds measured from lines-of-sight that are slightly offset from each other relative to the perpendicular to the ICME propagation direction are significantly affected by asymmetrical differences to the structure being sampled.
    \item Section \ref{sec:speed_stuff} : The layout for different lines-of-sight enables measuring whether or not the speed of the CME/ICME derived from structural characteristics with distance is different between different data sets (FR vs TEC vs {\em in-situ}), a potential source for physical insight.
    \item Section \ref{sec:speed_stuff} : Azimuthal offsets (from the `nose' of a CME/ICME) impact the speed observation made by {\em in-situ} spacecraft, increasing the variability. This effect is smoothed out with columnar integrations reducing the variability.
    \item Section \ref{sec:speed_stuff} : Remotely observing the CME/ICME with integrated lines-of-sight will show the structure when enough of the structure is coherent along the line-of-sight, reducing sensitivity to azimuthal differences in the CME/ICME path.
    \item Section \ref{sec:speed_stuff} : Differences in average speeds versus timing speeds from {\em in-situ} spacecraft measurements of ICME structure should be studied for those events with multi-spacecraft observations at similar offsets from the ICME propagation path.
    \item Section \ref{sec:speed_stuff} : Remote radio observations enable measuring different parts of the CME/ICME structure moving at different velocities from which an expansion rate can be estimated. {\em In-situ} measurements of ICMEs have shown that these structures can expand or contract crossing the spacecraft at different velocities between the leading and trailing edges.
    \item Section \ref{sec:speed_stuff} : A comparison of speeds observed {\em in-situ} timing versus TEC or RM structure indicate that the latter could be off by 2-17\%.
    \item Section \ref{sec:speed_stuff} : A technique using empirical scintillation properties from solar wind turbulence characteristics at 100 second integrations (a FETCH instrument parameter) should be studied for obtaining speeds. In combination with timing speeds and an analysis of actual multi-spacecraft ICME crossing data, a better understanding of ICME structure could be obtained.
\item Section \ref{sec:trailing_edge} : As shown in Table \ref{tab:selected times}, the Leading-Edge times for the magnetic flux rope of the ICME correspond closely between the Frequency Fluctuation technique and the {\em in-situ} analysis of simulated Sun-Earth spacecraft data.
    \item Section \ref{sec:trailing_edge} : Tomographic analysis can provide important information on the trailing edge location of a magnetic flux rope in a CME/ICME. This result is important for magnetic flux rope fitting, because the orientation and size are required to reproduce a FR measurement will change based on where the leading and trailing edges of the magnetic flux rope are placed. Properly measuring the size also impacts the magnetic field strength of the flux rope fit.
    \item Thorough analyses fusing Faraday rotation with other forms of remote radio data, white-light data, {\em in-situ} data, and modeling should be undertaken (e.g. \citealp[]{2010SoPh..265...49B}).
\end{itemize}

It is essential to acknowledge a critical factor when interpreting these findings: the challenge of accurately modeling the impact of MHD waves and unforeseen structures within CMEs/ICMEs on the measurements obtained. Pioneer `W' structures, as highlighted in studies by \cite{Jensen:2008} and \cite{Jensen:2018}, are an enigma. While the `W' structures likely have ties to CMEs/ICMEs, there are too few cases to be certain. The detection of structures trailing a CME/ICME suggests the presence of unknown dynamics, underscoring where reality diverges significantly from theoretical simulation outcomes. Despite our efforts to incorporate incoherent MHD wave noise (roughening) into these observations, our inability to generate a Kolmogorov spectrum underscores the complexity of the task at hand. Therefore, we advocate for caution in extrapolating from these results due to these inherent limitations. The presense of waves and unexpected plasma structures in Faraday rotation observations have only been studied by a few dedicated researchers, and much more work can be done when more data is collected.

Note that a new theory has been proposed to complement FR observing: the $B_\perp$-mode \citep{2024ApJ...963...25J}. This mode is sensitive to the component of the magnetic field {\em perpendicular} to the LOS but not its direction. When observed in combination with FR, the total magnetic field strength could be estimated. We do not develop how this observation would impact our analysis in this study, so we encourage the reader to keep this in mind with respect to our results. 

\section{Discussion and Conclusions}\label{conclusions}
We demonstrate that FETCH can remotely measure the magnetic field and TEC in a way that is needed to characterize the global 3-dimensional distribution of plasma within ICMEs between the Sun and Earth.
Every Faraday rotation experiment has compared its results to various models for the corona and heliosphere and structures within them. MHD modeling of the heliosphere has significantly advanced by incorporating multiple structures simultaneously, such as CMEs/ICMEs and corotating interactive regions. This is the first effort to utilize a realistic model, the AWSoM 3D MHD model, and analyze how remote sensing Faraday rotation, Total Electron Content, and speed from timing observations compare to the known simulated input as well as {\em in-situ} observations of the same structures.

We discovered that the asymmetry of these structures has a significant influence on the analysis results. For example, the view direction produced markedly different timing speeds. Despite these differences, they were less variable than {\em in-situ} speed estimates. Analyzing the Total Electron Content in the ICME was similarly impacted by viewing asymmetry, enabling a rough tomographic inversion for estimating density. 

The current approach to analyzing Faraday rotation observations is to initially assume that the magnetic field is calculated from RM/TEC*constant. This paper demonstrates that the assumption best applies where there are not significant variations in density and magnetic field. It's generally a factor of two off under normal conditions.

The timing of ICME structures crossing the different Lines between FETCH-hosting spacecraft enables observing the speed and expansion of the ICME. Detection of the sheath region is based on the TEC in the line of sight as it rises and falls, which also generates a signature in the FR, a measure that is weighted by the density. This is different from an {\em in-situ} spacecraft observation, which can be located in the path of the ICME or off to the side. 

The FR and TEC measurements are sensitive to different aspects of the structure, the sheath versus the flux rope versus the surrounding solar wind. Due to the asymmetry in these structures, timing when the largest coherent part of the structure is in the line-of-sight can vary significantly with just 15 degrees of angular offset. The implication is that tomography is a powerful tool to utilize these differences to determine the 3D spatial distribution.

We find that the leading edge to an ICME flux rope can be detected using frequency fluctuations for the steep decline in the inner edge of the sheath. Identifying the Trailing Edge to an ICME flux rope can be undertaken with a rough tomographic analysis. This pinpointing of the magnetic flux rope edges compared simulated {\em in-situ} spacecraft observations to the remote observing of TEC to make this determination. Further investigation with other examples of CMEs/ICME should be undertaken to inspect model specific differences. 

A critical aspect to being able to make a tomographic analysis was the placement of four radio frequency instruments on either side of the Earth-Sun line. Spacecraft radio Faraday Rotation measurements collected from the L5-L4 sensing path is a novel approach for studying Earth-bound ICMEs. We find that the FETCH instrument concept presents the opportunity to combine the study of ICME basic structure and physics with improved early-detection capabilities for high-risk geospace weather. 

Available codes and simulation results are released on Zenodo (\citealp{jensen_2024_10987687}).

\section{Acknowledgments}
We would like to thank Alexei Pevtsov at the National Solar Observatory, Lynn B. Wilson at NASA Goddard Space Flight Center, and Bernard V. Jackson at University of California San Diego for their feedback on these ideas. E.A. Jensen was supported by ACS Engineering \& Safety LLC and GSFC PHaSER Program with George Mason University for this project. Goddard Space Flight Center grant 80NSSC21K1991 supported the software purchases for this work.  L.K. Jian is supported by NASA's STEREO mission and HGI Grant 80NSSC23K0447. Basic research at the U.S. Naval Research Laboratory (NRL) is supported by 6.1 Base funding.  We gratefully acknowledge the supercomputing resources with which these simulations were performed: the Pleiades system provided by NASA’s High-End Computing Program under award SMD-11-2364, and the NCAR Yellowstone supercomputer. DeepAI was used to improve the language of this text at an earlier stage of development, and I have reviewed and thoroughly edited the results to make sure that it did not change the intent or meaning of any sentences.

\appendix
\section{Measuring Velocity/Speed Alternative} \label{app:instvel}
\begin{figure}[htb!]
	\begin{center}
	\includegraphics[width=\textwidth]{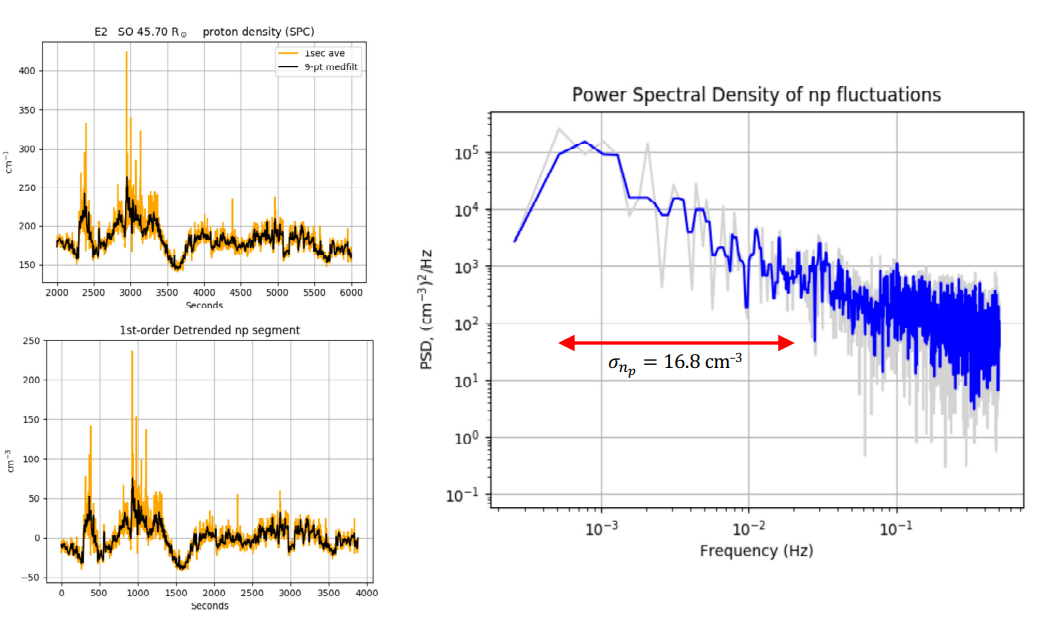}
	\caption{Calculation of the RMS proton density fluctuations. The upper left panel shows the proton number density detected by SWEAP SPC on 2019 April 1 during PSP's second perihelion encounter (E2) at a solar offset of $45.7\,\mathrm{R}_\odot$. The orange and black data are the 1-sec averaged and 9-point median-filtered data, respectively. The lower left panel shows the same data, detrended using a linear fit. The right panel shows the power spectral density, $PSD$, of the proton density fluctuations. The blue and gray data correspond to the 1-sec averaged and 9-point median-filtered data, respectively. Using a frequency range of $0.5-20\,\mathrm{mHz}$, the blue data yield an RMS fractional fluctuation of $\epsilon = 0.086$.}
	\label{fig:append}
	\end{center}
\end{figure}

One approach to estimating velocity/speed utilizes the turbulent spectrum that results from the scale sizes of density fluctuations. It is an empirical approach that has been utilized for scintillation studies. Ground-based radio wave scintillation has long been used to infer both the plasma density and velocity of CMEs/ICMEs with tomographic analysis providing the 3D maps of these quantities, for example see \citealp{Jackson:2004, Jackson:2006}. Scintillation is the twinkling in phase and intensity that a radio source undergoes due to changes in plasma density as it flows across the line-of-sight. CME/ICME impacts on scintillation continue to be new territory (example \citealp{2023SoPh..298...22T} and \citealp{2023SpWea..2103396M}). Among the challenges is that the effect on scintillation measures is small, a difficult measure for the low power sources that make up most IPS sources, a non-issue for many spacecraft applications. If this technique can be developed for FETCH, then scintillation-derived velocity/speed measures can be obtained rather than relying on timing studies between Lines 1-4 for speed estimates alone. Here we discuss the only model available for estimating this possibility.

To reiterate, this is new territory. Potential issues with this technique can include that the statistics for the CME/ICME plasma turbulence can be different from that for regular solar wind. Additionally, there's no reason to assume that the same turbulence statistics apply between CMEs/ICMEs. For example, \cite{2002GeoRL..29.1913L} investigated IPS measurements of a glancing CME crossing. They experienced issues with an enhancement that occurred in the intensity variance that could not be correlated with a density enhancement. They suggest that this enhancement is not due to velocity rather scale size of the turbulence; however, they were not analyzing the spectra for the first minimum in the Fresnel knee discussed by \cite{1990MNRAS.244..691M}. Additionally, this paper pre-dates \cite{Jensen:2018} which discussed the potential impact of reconnection in the signal, which would not have been included in their velocity model.

One model for studying the relationship between solar wind speed, $V_{SW}$, and fluctuations in electron density in spacecraft radio data was developed by \cite{Wexler:2019}. They started with the premise that radio frequency fluctuation, $\delta f(t)$, can be related to the temporal rate of change in electron density, $n_e$:
\begin{equation}
\delta f(t) = \frac{1}{2 \pi}r_e \lambda L_{LOS} \frac{d}{dt}n_e(t)
\end{equation}
with radio wavelength $\lambda$, $r_e$ is the classical electron radius and $L_{LOS}$ is the element integration length (commonly called the correlation scale of fluctuations). This formula gives the frequency fluctuation (FF) attributed to a single plasma slab of the given correlation length, passing across the radio line-on-sight (LOS). Considering the oscillations to be in the form of $e^{i\omega t}$ sinusoids, the expression may be carried into the frequency domain by Fourier transform, then presented in terms of power spectral density as a function of angular frequency, $\omega$.  Integrating over a frequency range of interest (usually the inertial range), the variance of density fluctuation is found as:
\begin{equation}
\sigma_{n_e}^2=\frac{1}{r_e^2L_{LOS}R}\int_{\nu_{low}}^{\nu_{high}}\frac{|FM(\nu)|^2}{\nu^2}d\nu
\end{equation}
where R is the solar offset to the point of closest solar approach along the radio LOS, and FM is the wavelength-normalized frequency fluctuation now given in Hz. The above expression already adds up all the slabs along the the effective integration length, set to R, such that the number of fluctuating density slabs is $R/L_{LOS}$ (see \cite{Wexler:2019}). Note that this equation does not give solar wind speed directly, but provides the needed estimate of $\sigma_{n_e}$ that will be used in specific models to estimate $V_{SW}$ (see below). The starting point for correlating observed radio frequency fluctuations to solar wind speed is therefore determination of $\sigma_{n_e}^2$. This cannot be done meaningfully for bulk heliospheric models, so we give an example using Parker Solar Probe observational data, as follows.

To develop an estimate for the variation in plasma density resulting from plasma density fluctuations, we examined data from the Solar Wind Electrons Alphas and Protons \citep[SWEAP][]{Kasper:2016} Solar Probe Cup (SPC) on board the Parker Solar Probe (PSP), corresponding to PSP's second perihelion encounter. We sampled data from 2019 April 1; although the full data set is available, e.g., in Figure 4 of \cite{Rouillard:2020}. The sample data set is 4000 sec ($\gtrsim1\,\mathrm{hr}$), with the proton plasma density, $n_\mathrm{p}$, averaged to a one-second cadence (upper left panel of Figure ~\ref{fig:append}). We chose this data because the solar offset was $45.7\,\mathrm{R}_\odot$, which is representative of the distances probed by the LOS between MOST 1 and 4 and between MOST 2 and 3.

We detrend the data using a linear fit, resulting in fluctuations about the zero-line (lower left panel of~\ref{fig:append}). We then determine the power spectral density ($PSD$, right panel of~\ref{fig:append}); the low frequency limit is determined by the data record length ($1/4000\,\mathrm{sec}=0.25\,\mathrm{mHz}$) and the high frequency limit is set by the Nyquist frequency corresponding to the data sampling rate ($0.5\times1/1\,\mathrm{sec}=500\,\mathrm{mHz}$). The power spectral density of $n_\mathrm{p}$ fluctuations exhibits a negative power scaling ($\alpha\approx-1.6$) from $1-20\,\mathrm{mHz}$ similar to the Kolmogorov spectrum ($\alpha\approx-5/3$) expected for MHD turbulence in the inertial regime. At high frequencies, a noise floor is reached at $\approx114\,(\mathrm{cm}^{-3})^2\,\mathrm{Hz}^{-1}$.

After subtracting the noise floor, the fluctuation variance can be determined by 
\begin{equation}
    \sigma_{n_\mathrm{p}}^2 = \int_{\nu_l}^{\nu_h} PSD(\nu)\mathrm{d}\nu
\end{equation}
where $\nu_l$ and $\nu_h$ are the low and high frequency limits, respectively, defining the negative power scaling range. The RMS fluctuations in $n_\mathrm{p}$ are then given by $\sigma_{n_\mathrm{p}}\equiv\sqrt{\sigma_{n_\mathrm{p}}^2}$. Over the range $0.5-20\,\mathrm{mHz}$ (the expected detectability range for FETCH), the RMS fluctuations for this sample is $\sigma_{n_\mathrm{p}}=16.8\,\mathrm{cm}^{-3}$. Dividing by the mean density over this sample, $n_\mathrm{p}=182.5\,\mathrm{cm}^{-3}$, we obtain a fractional fluctuation of $\epsilon\equiv\sigma_{n_\mathrm{p}}/n_\mathrm{p} = 0.086$.

We also computed the fractional fluctuation $\epsilon$ for the fourth PSP perihelion encounter over a range in data record lengths. For an outer scale of 2400 sec (0.4 mHz frequency limit), $\epsilon=0.07$; for a larger outer scale of 10800 sec (0.1 mHz frequency limit), $\epsilon=0.10$. For comparison, \cite{Wexler:2019} estimated $\epsilon=0.08$ at $\approx10\,\mathrm{R}_\odot$ using transcoronal radio data. Similarly, \cite{Krupar:2020} found $\epsilon=0.06-0.07$ in PSP data from perihelion encounters 1 and 2 at a temporal scale of 300 seconds. Note, however, much higher fractions may be expected in localized regions during transient solar wind phenomena.

\cite{Wexler2020} furthered the radio frequency fluctations analysis using X-band transcoronal recordings from the Akatsuki spacecraft near superior conjunction.  They noted that the density disturbances crossing the sensing LOS could be due to either acoustic waves if the regime close to the Sun where the wind speed ($V_{SW}$ is low, or due to advected density fluctuations in the regime with SW speed $\gg$ sound speed. The model proposed for the latter case, which is of interest here, is based on: 
\begin{equation}
    \frac{d}{dt}=V_{SW}\cdot \nabla
\end{equation}
where $\nabla$ is enacted as the reciprocal of characteristic length scale of density variation moving radially across the LOS. Then the time derivative expressed as the equivalent frequency is $V_{SW}/L_{RAD}$ and the RMS wavelength-normalized frequency fluctuation becomes:
\begin{equation}
    \sigma_{FM} = r_e\frac{V_{SW}}{L_{RAD}}\epsilon n_e \sqrt{L_{LOS}R}
\end{equation}
from which the solar wind outflow speed can be estimated. 

Application of this method to the study of CMEs crossing radio lines-of-sight won't be trivial.  In addition to obtaining satisfactory measurements of $\sigma_{FM}$, realistic estimates of the electron density, fractional density fluctuation and correlation length (representative slab thickness) will be needed. Further, the method requires an estimate of the spatial scale of density fluctuation, $L_{RAD}$ in the wind moving across the sensing path. An appropriate integration length will need to be established; for that, high-caliber MHD models should be most useful.

\section{Tomographic Approach} \label{app:tomo}
The tomographic calculation was undertaken by aligning the time series of lines 1-4 to coincide with a time stretch and time shift function, such that each observation of the sheath region was set to the same baseline time. The baseline was set by both lines 2 and 3. When density is calculated, which azimuthal offset region it is in can be determined, for example Region C versus Region B in Figure \ref{fig:linescomp}.

A complication was present to the layout of the regions as shown in Figure \ref{fig:linescomp} with the presence of a stream interaction region (see `CIR' in Figure \ref{fig:Line1}, Panel L1). Utilizing densities observed by the spacecraft immersed in the SIR will require either shifting the region boundaries or removing the SIR contribution. We selected the latter option.

\begin{figure}[htb!]
	\begin{center}
	\includegraphics[width=2.9in]{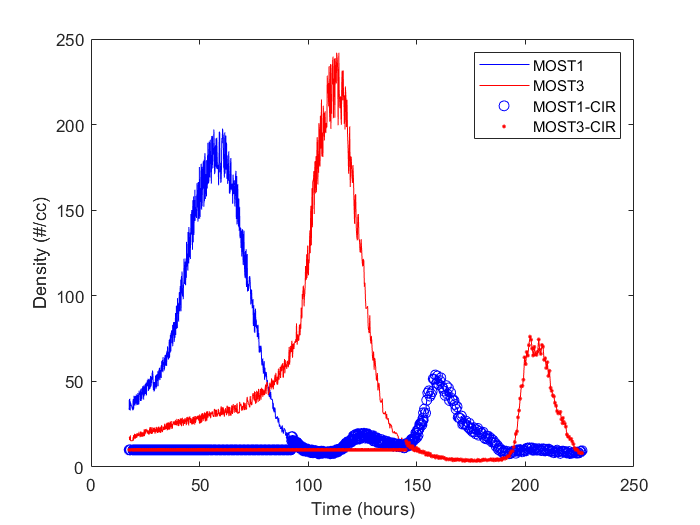}
	\includegraphics[width=2.9in]{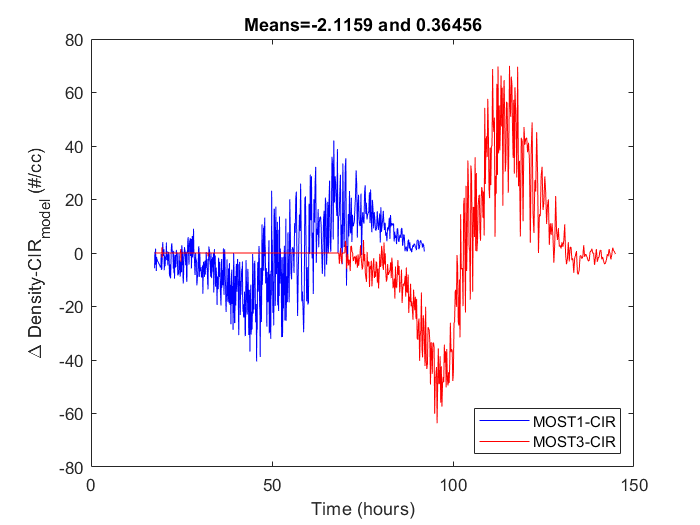}
	\includegraphics[width=2.9in]{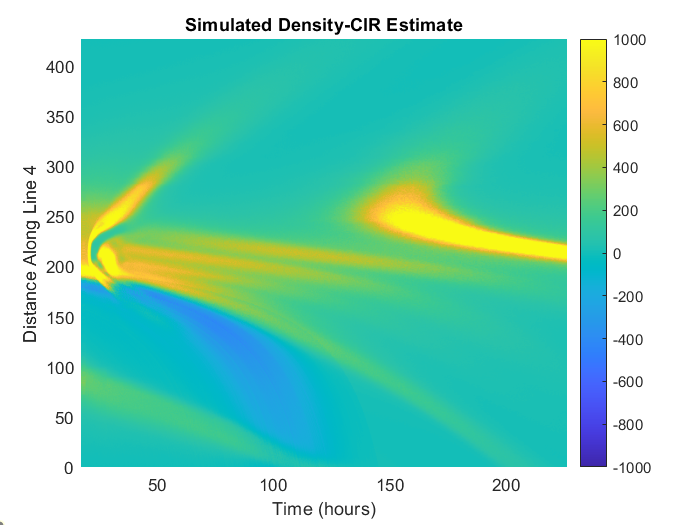}
	\includegraphics[width=2.9in]{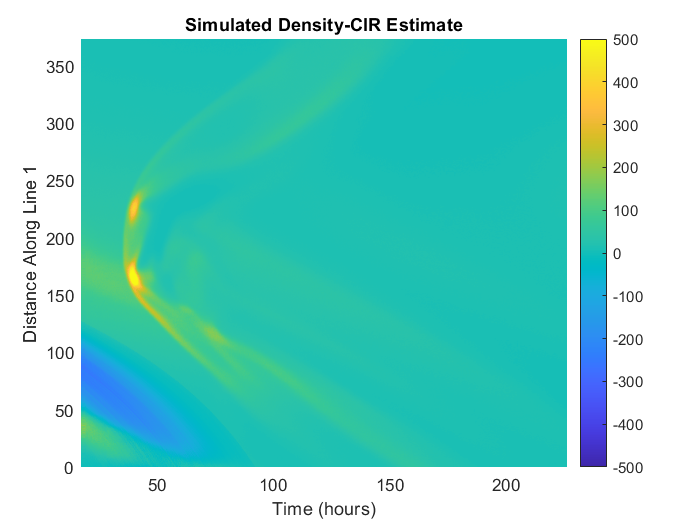}
	\caption{Approximate SIR/CIR removal. The top left panel shows the adjusted densities at MOST1 and MOST3 from the CIR removal procedure; the density is set for 10 $m^{-3}$. The top right panel shows the remainder after the density model is applied. The bottom panels show how the CIR model, constructed based on timing at MOST1 and MOST3, compares to the AWSoM input.}
	\label{fig:CIRmod}
	\end{center}
\end{figure}

For detailed information on how the FETCH spacecraft layout would detect corotating/stream interaction regions (SIRs/CIRs), see \cite{2023ApJ...955...90W}. To remove the SIR/CIR, the steps were to determine the boundaries in the MOST1 (13/1700-16/2200 hours) and MOST3 (13/1700-19/0100 hours) time series. Note that the time series starts on 2005 May 13 at 1700. This time frame was used to calculate the rate of rotation of the SIR. Figure \ref{fig:CIRmod} (left side) shows the time series for each spacecraft before and after SIR removal. An $r^{-1}$ scale falloff was necessary to prevent the TECs from going negative. Figure \ref{fig:CIRmod} (right side) illustrates that (1) the densities along Lines 1-4 were cumulatively similar with the adjusted falloff, but (2) the CIR was rotating at varying rates relative to its density distribution. While the plots shown give the simulation output, these issues demonstrate the strength of the FETCH layout to address solar wind dynamics. By applying a fixed rate of rotation from MOST1 to MOST3, the position of the CIR arrived at the correct time, but its curvature with offset from the Sun is obviously not radial in the simulation (the assumption applied in this removal process). As CIRs can have different curvatures, we continued with this radial model to see if having the correct CIR model mattered to the overall results. Ultimately, the TEC along the LOS will have the TEC from the CIR subtracted. Where the CIR is located is not a significant issue in the TEC calculation. Why we had to use the $r^{-1}$ falloff could be related to this displacement of the CIR issue. It is in the following paper where the CIR and the heliospace in general will be modeled with an Archimedian curvature.

With the TEC contribution from the SIR/CIR removed, the adjusted TECs for the 4 lines-of-sight was then stretched relative to time. The arrivals of the maximum sheath density and the minimum fluxrope density on lines 2 and 3 were averaged, taken as the model for the time stretches on lines 1 and 4. Then the adjusted TECs were shifted in time to align with lines 2 and 3. Next, all the time-series were resampled to even out the time steps, which were irregular. The time periods analyzed are cut off after the passage of the ICME, because another SIR/CIR entered the lines of sight. Finally, the known regions as shown in Figure \ref{fig:linescomp} are subtracted out from the adjusted TECs. This resulted in some negative values on lines 1 (before the ICME) and 3 (after the ICME), which were set to 0.1 to flag an issue with the particular time.

\begin{figure}[htb!]
	\begin{center}
	\includegraphics[width=6in]{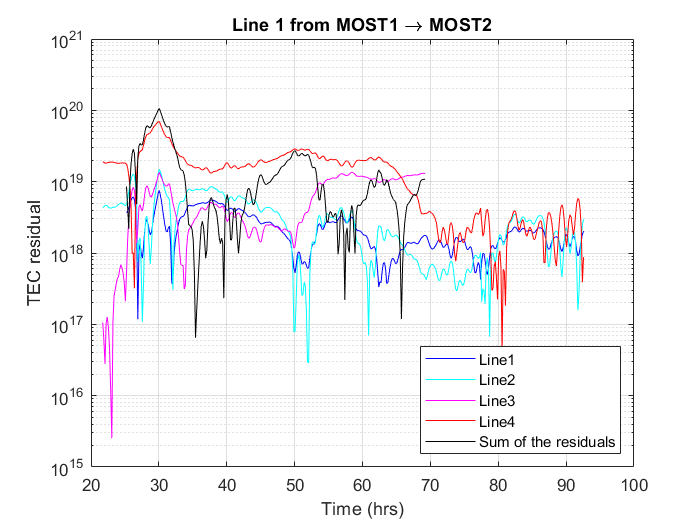}
	\caption{Errors in TEC from the density solutions. The difference between the initial TEC values and the $\Psi\eta$ values are shown. Line 4 errors dominate showing that the ICME evolved significantly from it to the other lines. The black curve provides information on the direction of the errors; when it is lower than Line 4, the other residuals are negative. Flagged TEC values are removed from this plot.}
	\label{fig:matrixerr}
	\end{center}
\end{figure}

The steps taken above detail how the TECs on each line of sight were aligned in time for the structure of the ICME to cross each at the same time. This is to enable a snapshot of the density of the structure in the unknown regions shown in Figure \ref{fig:linescomp}. At one point in time, the same density is crossing each line, with the radial falloff comprising the difference between results. Ideally, for each point in time then there are 4 adjusted TECs and 4 unknown regions. The following calculation can then be made.

\begin{equation}
\begin{bmatrix}
\Psi_{A1} & \Psi_{B1} & \Psi_{C1} & \Psi_{D1}\\
\Psi_{A2} & \Psi_{B2} & \Psi_{C2} & \Psi_{D2}\\
\Psi_{A3} & \Psi_{B3} & \Psi_{C3} & \Psi_{D3}\\
\Psi_{A4} & \Psi_{B4} & \Psi_{C4} & \Psi_{D4}
\end{bmatrix}
\begin{bmatrix}
\eta_{A}\\
\eta_{B}\\
\eta_{C}\\
\eta_{D}
\end{bmatrix}
=
\begin{bmatrix}
TEC_{Line1}\\
TEC_{Line2}\\
TEC_{Line3}\\
TEC_{Line4}
\end{bmatrix}
\end{equation}

\begin{equation}
    [\eta]=Inverse([\Psi])[TEC]
    \label{eqn:4eq4unk}
\end{equation}

The $[\Psi]$ matrix is composed of the following assumptions. The TEC segment $\int_i^f{N\left(\frac{R_{ref}}{r(S)}\right)^2dS}=\eta\int_i^f{\left(\frac{R_{ref}}{r(S)}\right)^2dS}=\Psi\eta$. The subscripts for $\Psi$ specifically refer to the region from Figure \ref{fig:linescomp} and the number for the line of sight. When $N$ is assumed fixed after accounting for the location of the step along $dS$, it's been relabeled as $\eta$.

The calculation shown in Equation \ref{eqn:4eq4unk}, has an issue in that the solution is not bound to keep density positive. As a result, the output included negative values. To examine if the result contains useful information, it was first adjusted by shifting all curves upward to positive and then reducing the magnitude. This produced significant errors. Then the absolute value of the result was calculated. This also produced significant errors; however, the structure of the input ICME became visible. Analysis of the errors indicated that the calculation in Equation \ref{eqn:4eq4unk} required adjustment for $\eta$ depending on which line the tomography was to be compared to. The calculation for $\Psi$ that we constructed used line 1 for the reference. To adjust the $\eta$ solutions, these were divided by the difference in the different region lengths squared. The solution shown in Figure \ref{fig:tomogL2} has very similar densities to the input from the AWSoM model.

Finally, the performance of the model was inspected in Figure \ref{fig:matrixerr}. The densities calculated were consistently too low for Line 4 indicating that the ICME evolved significantly between the first crossing and later crossings. This was a weakness to the approach assuming a consistent structure.

\bibliography{FETCHcme}
\bibliographystyle{aasjournal}

\end{document}